\documentclass[reprint,
superscriptaddress,
amsmath,amssymb,
aps,
pra,
]{revtex4-2}
\usepackage{graphicx}
\usepackage{dcolumn}
\usepackage{bm}
\usepackage{dsfont}
\usepackage{physics}
\usepackage{color}

\begin{document} 

\title{Quantum Dynamics of Vibrational Polariton Chemistry}
\author{Lachlan P. Lindoy}%
\affiliation{Department of Chemistry, Columbia University, 3000 Broadway, New York, New York, 10027,  U.S.A}
\author{Arkajit Mandal}%
\affiliation{Department of Chemistry, Columbia University, 3000 Broadway, New York, New York, 10027,  U.S.A}
\author{David R. Reichman}
\email{drr2103@columbia.edu}
\affiliation{Department of Chemistry, Columbia University, 3000 Broadway, New York, New York, 10027,  U.S.A}

\begin{abstract}
We employ an exact quantum mechanical simulation technique to investigate a model of cavity-modified chemical reactions in the condensed phase.  The model contains the coupling of the reaction coordinate to a generic solvent, cavity coupling to either the reaction coordinate or a non-reactive mode, and the coupling of the cavity to lossy modes.  Thus, many of the most important features needed for realistic modeling of the cavity modification of chemical reactions are included.  We find that when a molecule is coupled to an optical cavity it is essential to treat the problem quantum mechanically in order to obtain a quantitative account of alterations to reactivity.  We find sizable and sharp changes in the rate constant that are associated with quantum mechanical state splittings and resonances.  The features that emerge from our simulations are closer to those observed in experiments than are previous calculations, even for realistically small values of coupling and cavity loss.  This work highlights the importance of a fully quantum treatment of vibrational polariton chemistry.
\end{abstract}

\maketitle
\section{Introduction}

A series of recent experiments~\cite{ThomasACID2016,ThomasS2019,LatherACID2019,VergauweACID2019,AnoopNp2020, LatherCS2022, SimpkinsJPCC2021, NagarajanJACS2021} have suggested that when molecular vibrations are coupled to the radiation modes inside an optical cavity, ground state chemical kinetics can be both enhanced~\cite{ThomasACID2016, ThomasS2019, NagarajanJACS2021} or suppressed~\cite{LatherCS2022,LatherACID2019, NagarajanJACS2021}.  Such effects are purported to operate in the absence of external optical pumping~\cite{KenaACSCS2019} and have been hypothesized to arise from the hybridization of molecular vibrational states and the photon (Fock) states of a cavity radiation mode~\cite{ThomasS2019, ThomasACID2016}. The interpretation of these experiments is still a matter of debate, and thus the viability of marked alterations in chemical reactivity remains an open topic.  Indeed, while the spectroscopic fingerprints of light-matter hybridization such as the Rabi-splitting observed in the IR spectra are manifest, the significance of the coupling to radiation modes for markedly changing chemical reactivity is unclear.

Theoretical studies that describe both the radiation modes as well as the molecular vibrations using classical mechanics have had limited success in describing currently available experiments~\cite{LiNC2021, LiJPCL2021, MandalJCP2022, LindoyJPCL2022, LiJCP2020, JorgeJCP2020, SunJPCL2022, WangJPCL2022, ChristianArxiv2022}.  Specifically, previous work~\cite{LiNC2021, LiJPCL2021, MandalJCP2022} using the Grote-Hynes (GH) rate theory~\cite{GroteJCP1980, PeterRMP1990, MelnikovJCP1986}, applicable in the limit of strong molecule-bath interactions, do show cavity frequency-dependent chemical kinetics modification. However these studies~\cite{LiNC2021, LiJPCL2021, MandalJCP2022} predict that the chemical reaction rate is  suppressed most strongly when the cavity frequency is near to the {\em barrier} frequency as opposed to {\em molecular vibrational} frequencies as seen in experiments, and that the rate profile is only weakly modified in an  extremely broad manner with respect to the cavity frequency $\omega_\mathrm{c}$, spanning thousands of wavenumbers ($\sim$ 5000 cm$^{-1}$). This is in stark contrast to experimental observations, where the width of the changes in the rate profile are on the order of $\sim$ 100 cm$^{-1}$~\cite{ThomasACID2016,ThomasS2019,LatherACID2019,VergauweACID2019,AnoopNp2020, LatherCS2022}. On the other hand, a recent study~\cite{LindoyJPCL2022} using the Pollak-Grabert-H\"anggi (PGH) theory~\cite{PollakJCP1989}, as well as direct trajectory-based computational work~\cite{SunJPCL2022,PhilbinJPCC2022}, have predicted {\it enhancement} of chemical rates when the molecule-bath coupling is relatively weak. Interestingly, these studies have predicted a significantly sharper rate profile than that which emerges at strong coupling, have demonstrated that the effect is more sizable, and have revealed that the chemical rate is most strongly enhanced when the cavity frequency is close to the reactant vibrational frequency~\cite{LindoyJPCL2022,SunJPCL2022,PhilbinJPCC2022}. While these studies move theory closer to laboratory observations, there is still a substantial gulf between experiments and our theoretical understanding.

A major missing component in the theoretical work discussed above is the quantum nature of problem. Thus, a direct account of even the formation of Rabi-split polaritonic states is omitted. Simple quantum corrections to the GH theory, such as found using quantum transition state theory~\cite{YangJPCL2021} or zero-point energy corrections to the energy barrier~\cite{FischerJCP2022} have been carried out, but these approximate calculations diverge from experimental expectations even more than their fully classical counterparts, showing, for example, an even broader range of alteration of the rate profile than that seen in classical calculations.  Recent fully quantum dynamical studies which ignore the explicit interactions of the molecule with the solvent degrees of freedom also do not find a resonant structure in the cavity frequency dependence of chemical rate~\cite{GalegoPRX2019, FischerJCP2022}. Taken as a whole, these studies point to the clear pressing need to perform exact quantum calculations on models that include the relevant molecular, solvent, and cavity degrees of freedom.

In this work, we use a customized version of the hierarchical equations of motion (HEOM) approach~\cite{TanimuraJPSP1989} (see Methods for further details) to {\em exactly} simulate cavity modified chemical kinetics of a single molecule coupled to a radiation mode as well as dissipative  molecular and solvent modes. The need to modify how the HEOM calculations are carried out is crucial for obtaining converged exact quantum dynamical rates.  As in some recent studies~\cite{PhilbinJPCC2022,MandalJCP2022}, we also include the coupling of the cavity mode to a bath which mimics cavity loss~\footnote{Note that the degree of cavity loss in our work is substantially smaller than that in recent classical calculations~\cite{PhilbinJPCC2022}.  Furthermore, even with these lower levels of cavity loss, the effects of loss are significantly more substantial than those found in recent theoretical work based on a classical mechanical description~\cite{PhilbinJPCC2022,MandalJCP2022}. Indeed, for the range of cavity loss used in this work, these previous studies would suggest that cavity loss has little to no  effects on chemical reactivity~\cite{PhilbinJPCC2022,MandalJCP2022}. The importance of quantum effects in this regard are explained in the main text}.  We show that coupling molecular vibrations to a cavity radiation mode can both enhance or suppress chemical reactivity, with the largest effect occurring when the cavity mode is near resonant with specific molecular vibrational modes.  Crucially, we find that the cavity frequency-dependent rate shows a much sharper profile ($\sim 100$ cm$^{-1}$) compared to what is predicted from classical rate theories~\cite{LiNC2021, LindoyJPCL2022} or with quantum corrections~\cite{WangJPCL2022, FischerJCP2022}, even when compared to recent theoretical results in the classical weak-coupling regime~\cite{PhilbinJPCC2022, SunJPCL2022}.  Our results also demonstrate that the details of the solvent-molecule interactions are extremely important, and that even realistically small rates of cavity loss can play a crucial role in enabling cavity modification of chemical reactivity. 

Specifically, we find that the extent and the nature (enhancement or suppression) of the cavity modification to the reaction rate depends sensitively on (a) the details of the molecular system, such as the potential energy surface and the vibrational eigenspectrum, (b) the details of the solvent, as encoded in the spectral density and solvent friction and (c) the details of the cavity (e.g. the light-matter coupling strength and the rate of cavity loss) and how it couples to matter. Our results reveal that the cavity modification of chemical rates can largely be rationalized by considering how molecular vibrational states are altered by hybridization with cavity photon states (forming so-called vibrational polaritons) to effectively increase or decrease the interaction of the molecule with its environment. The resonant structure in the cavity-modified reaction rate naturally arises from the hybridization of light and matter, which occurs most strongly when the cavity and matter states are in resonance. {\em These effects emerge from fundamental quantum light-matter interactions and cannot be fully captured with simple classical or semiclassical descriptions of light and matter.} 

In essentially all previous experimental work~\cite{ThomasACID2016,ThomasS2019,LatherACID2019,VergauweACID2019,AnoopNp2020, LatherCS2022, SimpkinsJPCC2021, NagarajanJACS2021} a large ensemble of molecules ($\sim 10^{10}$ molecule per cavity mode~\cite{JorgeNC2019,JavierNJP2015}) are collectively coupled to the cavity radiation modes. In contrast, most theoretical work, including the calculations we present here, operate in the single molecule limit and do not address collective effects in a
direct manner. Despite some studies addressing collective polaritonic behavior~\cite{MandalJCP2022, KansanenArxiv2022, DerekArxiv2022}, a detailed theoretical explanation for such
collective effects remains elusive. We would like to point out, however, that single molecule studies are possible in principle.  In this regard, as will be discussed below, it is crucial to note that our quantum calculations suggest that alterations to reaction rates may be observed with cavity coupling strengths that are orders of magnitude smaller than suggested in other recent single molecule studies.  This fact highlights the potential feasibility of experimental observation of modification of rates even in the single molecule limit.


\begin{figure*}
\centering
\includegraphics[width=1.0\linewidth]{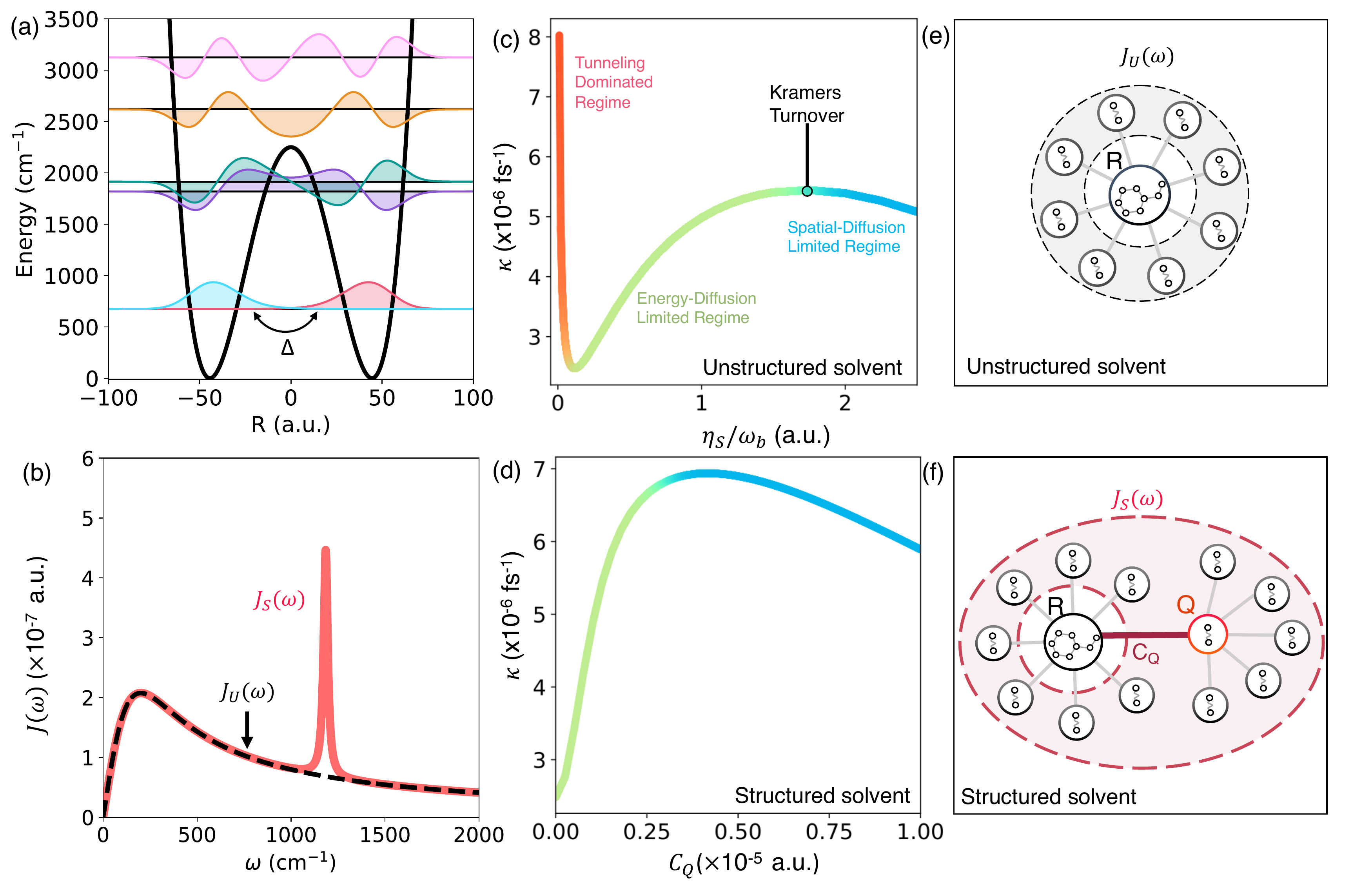}
\caption{\footnotesize \textbf{Effect of molecule-solvent coupling on chemical kinetics.} ({a}) Potential energy surface and vibrational eigenstates of model molecular system. ({b}) Effective spectral density for unstructured ($J_U(\omega)$, black dashed line) and structured ($J_S(\omega)$, red solid line) solvent environment as schematically depicted in (e) and (f), respectively. The peak at 1200 cm$^{-1}$ in $J_S(\omega)$ arises due strong coupling ($C_Q$) between the molecule and a spectator mode $Q$. (c) Chemical rate constant $\kappa$ as a function of solvent friction  $\eta_\mathrm{s}$ when the molecule is embedded in an unstructured environment, as illustrated in (e). Chemical rate constant $\kappa$ as a function of molecule-spectator mode coupling $C_Q$ when the molecule is embedded in an structured environment as illustrated in (f)}
\label{fig1}
\end{figure*}

\begin{figure*}
\centering
\includegraphics[width=0.99\linewidth]{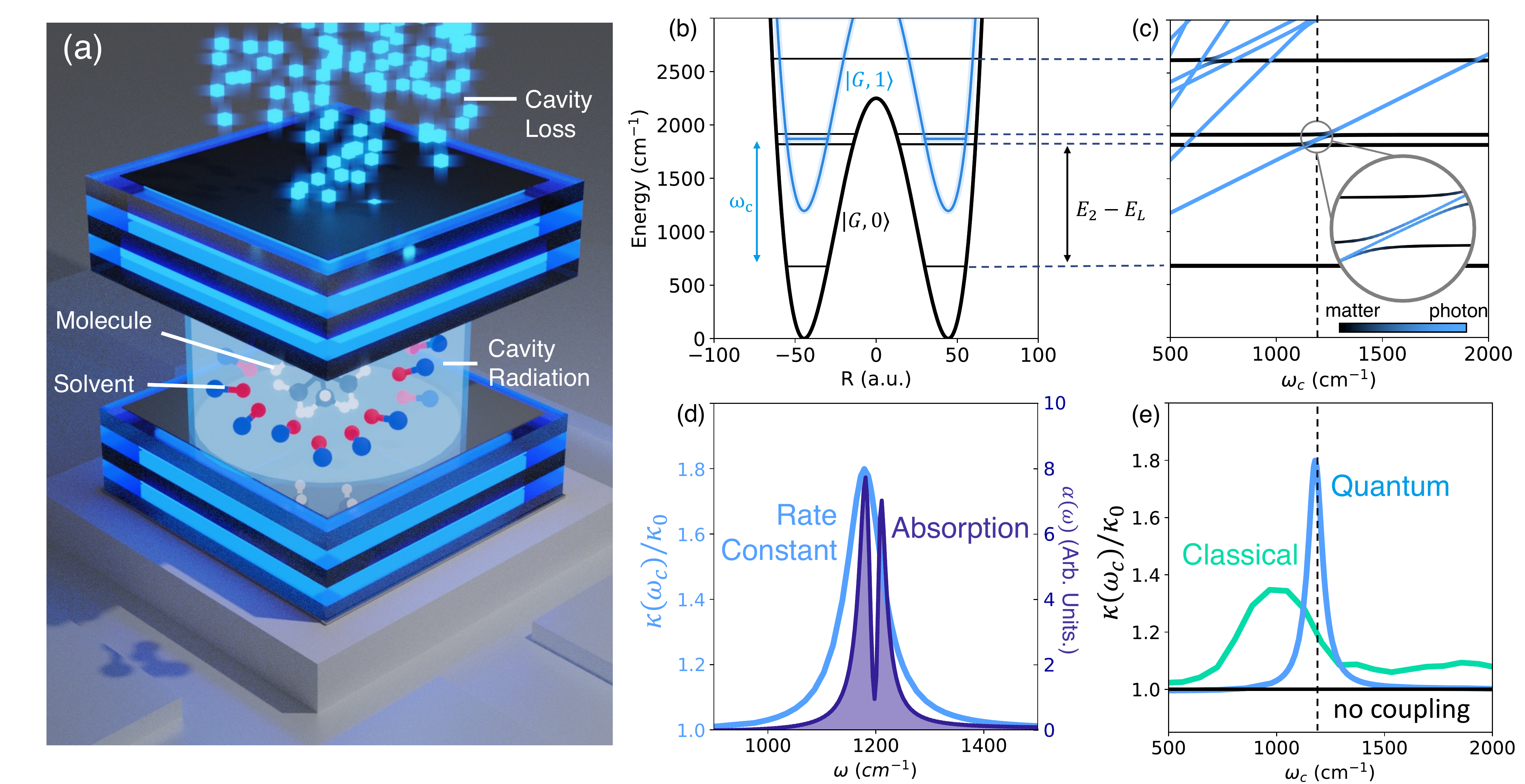}
\caption{\footnotesize \textbf{Resonant cavity modification of ground state chemical kinetics.} (\textbf{a})   Schematic illustration of a molecule coupled to a lossy cavity radiation mode as well as to other solvent molecules. (\textbf{b}) Potential energy surfaces for a molecular ground adiabatic state with 0 photons $|G,0\rangle$ (black solid line) and with 1 photon (blue solid line) $|G,1\rangle$, as well as the corresponding vibrational eigenstates of  $|G,0\rangle$ and $|G,1\rangle$ (horizontal sold lines). (\textbf{c}) Vibrational polariton eigenspectrum as a function of cavity photon frequency $\omega_c$.   (\textbf{d}) Absorption spectrum (violet shaded area) showing a polaritonic Rabi-splitting of $\sim$ 30 cm$^{-1}$ and the cavity modified rate constant $\kappa$ (normalized with the rate constant $\kappa_0$ outside of the cavity)  with a cavity lifetime $\tau_c=100$ fs (blue solid line). The linewidth of the absorption spectrum and chemical rate constant are similar in magnitude. (\textbf{e}) Comparing the cavity modified chemical rate constant computed using exact quantum (blue solid line) and classical (green solid line) dynamical simulations, showing the failure of classical description in quantitatively capturing the effects of quantum light-matter interactions. Here, the light-matter coupling is $\eta_c = 0.00125$  and solvent friction $\eta_s = 0.1\omega_b$.}
\label{fig2}
\end{figure*}

\section{Results} 
{\bf  Theoretical Model.}
The model quantum electrodynamics (QED) Hamiltonian used in this work is based on the Pauli-Fierz (PF) light-matter Hamiltonian in the dipole gauge in the single mode and long-wavelength limits, and is written as~\cite{Flick2017PNAS, MandalJPCB2020, Rokaj2018JPBAMOP}
\begin{align}
\hat{H} &= \hat{H}_\mathrm{mol} +\hat{H}_\mathrm{solv} + \hat{H}_\mathrm{cav} +   \hat{H}_\mathrm{loss},
\end{align}
where $\hat{H}_\mathrm{mol}$ is the molecular Hamiltonian, $\hat{H}_\mathrm{solv}$ describes solvent as well as molecule-solvent interactions, $\hat{H}_\mathrm{cav}$ is the cavity Hamiltonian describing a radiation mode and its interaction to matter in the dipole gauge, and $\hat{H}_\mathrm{loss}$ describes the cavity loss term.  

In this work we consider a molecular Hamiltonian $\hat{H}_\mathrm{mol} =  \hat{T}_{R} + V(\hat{R})$ that contains a one-dimensional reaction coordinate $R$. The ground state potential energy surface along this reaction coordinate,  $V(\hat{R}) =  \frac{\omega_b^4}{16E_b} \cdot \hat{R}^4 - \frac{1}{2}\omega_b^2\cdot \hat{R}^2 - c\cdot R^3$, takes the form of a double well potential.  In the main text we consider a barrier frequency $\omega_b = 1000$ cm$^{-1}$, barrier $E_b = 2250$ cm$^{-1}$, and a symmetric double potential with $c = 0$ (see Supplementary Note 2 for results with $c \ne 0$), as shown in Fig.~\ref{fig1}a (black solid line). The molecular Hamiltonian $\hat{H}_\mathrm{mol}$ can be equivalently represented using the vibrational states,

\begin{align}\label{Hmol}
\hat{H}_\mathrm{mol} &= \sum_i E_{i}|v_{i}\rangle\langle v_{i}|  \nonumber\\
&\equiv \Bar{ E}_0  \Big(|v_{R}\rangle \langle v_{R} |  +  |v_{L}\rangle \langle v_{L}| \Big) + \sum_{i \ge  2} E_{i}|v_{i}\rangle\langle v_{i}|  \nonumber \\
&+ \Delta \Big(|v_{R}\rangle \langle v_{L}| + |v_{L}\rangle \langle v_{R}|\Big), 
\end{align}
where $\{|v_{i}\rangle\}$ are the vibrational eigenstates of the molecular Hamiltonian ($\hat{H}_\mathrm{mol}|v_{i}\rangle = E_{i} |v_{i}\rangle$). In the second line we have introduced localized states $|v_{L}\rangle = \frac{1}{\sqrt{2}}(|v_{0}\rangle + |v_{1}\rangle)$ and $|v_{R}\rangle = \frac{1}{\sqrt{2}}(|v_{0}\rangle - |v_{1}\rangle)$, with an energy $\Bar{ E}_0 = \frac{1}{2}(E_0 + E_1)$ and a coupling $\Delta = \frac{1}{2}(E_{1} - E_{0})$. These states are the localized ground states of the left and the right wells (blue and red wavefunctions in Fig.\ref{fig1}a), respectively. We define the well frequency $\omega_0 = E_{2} - \Bar{ E}_0 \approx 1140$ cm$^{-1}$. 

$\hat{H}_\mathrm{solv}$ is taken as
\begin{align}
\hat{H}_\mathrm{solv} &= \frac{P_Q^2}{2} + \frac{1}{2}\omega_{Q}^2 \Big(Q + \frac{C_{Q} R}{\omega_Q^2}\Big)^2  \nonumber \\
&+ \sum_j \frac{\hat{P}_{j}^2}{2} + \frac{1}{2}\Omega_{j}^2 \Big(\hat{X}_j + \frac{C_{j} \hat{R}}{\Omega_j^2}\Big)^2 \nonumber \\
&+ \sum_j \frac{\hat{p}_{j}^2}{2} + \frac{1}{2} {\omega}^2_{j} \Big(\hat{x}_j + \frac{{c}_{j} \hat{Q}}{{\omega}^{2}_j}\Big)^2.
\end{align}
The first line describes a spectator mode with coordinate $Q$ (or equivalently a collective solvent mode~\cite{HughesJCP2009, GargJCP1985, MandalJCP2022}) coupled to the reaction coordinate. The second line describes a set of dissipative solvent modes described by a broad spectral density $J_{U}(\Omega) =\frac{\pi}{2} \sum_{j} \frac{ {C}_j^2}{ {\Omega}_j} \delta(\Omega -{\Omega}_j) = 2{\Lambda}_{s}\Omega{\Gamma}/(\Omega^2+ {\Gamma}^2) = {\eta}_{s}\Omega{\Gamma}^2/(\Omega^2+ {\Gamma}^2) $ (black dashed line in Fig.~\ref{fig1}a).  The third line describes a set of secondary solvent modes $\hat{X}_j$ that couple to the spectator mode coordinate $Q$ and are also described with a broad spectral density ${J}_{u}(\omega) = \frac{\pi}{2} \sum_{j} \frac{c_j^2}{\omega_j} \delta(\omega -\omega_j) = 2\lambda_{s}\omega\gamma/(\omega^2+\gamma^2)$.   

The cavity Hamiltonian $\hat{H}_\mathrm{cav}$ describing a single cavity mode and its coupling to matter is given by 
\begin{align}
\hat{H}_\mathrm{cav} &= \frac{\hat{p}_\mathrm{c}^{2}}{2} + \frac{1}{2} \omega_\mathrm{c}^2 \Big( \hat{q}_\mathrm{c} + \sqrt{\frac{2}{\omega_{c}}} \eta_c \cdot \hat{\mu}  \Big)^2.
\end{align}
Here, $\omega_\mathrm{c}$ is the cavity photon frequency and $\eta_c = \frac{1}{\omega_\mathrm{c}}\sqrt{\frac{\hbar \omega_\mathrm{c}}{2\epsilon_{0} V}}\hat{e}$ is the light-matter coupling strength with vacuum permittivity $\epsilon_{0}$, quantization volume $V$ and the direction of polarization $\hat{e}$. Further, $\hat{p}_\mathrm{c} = i\sqrt{\frac{\hbar\omega_\mathrm{c}}{2}} (\hat{a}^\dagger - \hat{a})$ and   $\hat{q}_\mathrm{c} = \sqrt{\frac{\hbar}{2\omega_\mathrm{c}}} (\hat{a}^\dagger + \hat{a})$,  where $\hat{a}^\dagger$  and $\hat{a}$ are the photon creation and annihilation operators,and $\hat{\mu}$ is the matter dipole operator. 

Finally $\hat{H}_\mathrm{loss}$ describes the bath that is coupled to the cavity mode which enables cavity loss, 

\begin{align}
\hat{H}_\mathrm{loss} = \sum_k \frac{\hat{\Pi}_{k}^{2}}{2} + \frac{1}{2}\tilde{\omega}_{k}^{2} \Big(\hat{\mathcal{Q}}_k + \frac{\mathcal{C}_{k} \hat{q}_c}{\tilde{\omega}_k^{2}}\Big)^2,
\end{align}
where $\tilde{\omega}_{k}$ and $\mathcal{C}_{k}$ which control cavity leakage are described via the spectral density
${J}_{L}(\omega) = \frac{\pi}{2} \sum_{k} \frac{\mathcal{C}_{k}^2}{\tilde{\omega}_{k}} \delta(\omega -\tilde{\omega}_{k}) = 2\lambda_{L}\omega\gamma_{L}/(\omega^2+\gamma_{L}^2) $.  With this spectral density, the cavity loss rate is defined as $\Gamma_{c} = 1/\tau_c = 2 J(\omega_{c})/(1- e^{-\beta\omega_c}   )$ where $\tau_c$ is the cavity lifetime and $\beta = 1/k_{B}T$ with the Boltzmann constant $k_{B}$ and temperature $T$. Physically, our model Hamiltonian flexibly contains nearly all essential ingredients assumed to influence chemical reactions in a cavity. Further details of the model parameters are provided in the  Supplementary Note 1.

{\bf  Chemical Kinetics Outside a Cavity.} The chemical kinetics of a  molecular system in the absence of the cavity (setting $\eta_c = 0$) embedded in an unstructured solvent (setting $C_Q = 0$) depends on the molecule-solvent interaction strength. Such a setup is schematically illustrated in Fig.~\ref{fig1}e and the corresponding bath spectral density $J_{U}(\omega)$ is presented in Fig.~\ref{fig1}b (black dashed line). The chemical rate constant $\kappa$ (see Methods), obtained from exact quantum dynamics simulation using a specialized HEOM approach (see Methods) as a function of $\eta_s$ is presented in  Fig.~\ref{fig1}b, and shows three distinct regimes. For very low $\eta_s$ the chemical kinetics is dominated by direct nuclear tunneling, that is the transition $|v_L\rangle \rightarrow   |v_R\rangle$ via the tunnel-coupling $\Delta$ in Eq.~\ref{Hmol}.  We refer to this regime as the tunneling-dominated regime. In this regime, an increase in the molecule-solvent interaction (by increasing $\eta_s$) leads to a sharp decline in the reaction rate~\cite{QiangJCP2011} as the bath (solvent) degrees of freedom effectively renormalize and lower $\Delta$.  

While the increase in $\eta_s$ reduces the direct nuclear tunneling, an alternate reaction pathway involving thermal excitations, that is $|v_L\rangle \rightarrow \{|v_i\rangle\}\rightarrow |v_R\rangle$, starts to play an increasingly important role. For $\eta_s>0.1\omega_{b}$, the later pathway becomes the dominant one. For $1.8\omega_{b}>\eta_s>0.1\omega_{b}$, the overall reaction rate is limited by the equilibration rate of the vibrational states in the left well. In this regime, akin to the energy diffusion-limited regime in the Kramers turnover problem~\cite{PeterRMP1990} \footnote{Throughout this work we will refer to the peak in rate vs coupling plot as the Kramers turnover point, and, interchangeably, the regime before the peak as the weak-coupling or ``energy diffusion-limited''  regime and that after the peak as the strong-coupling or ``spatial diffusion-limited'' regime. It should be noted that when the reaction rate is controlled by multiple parameters, the use of the terms such as ``energy diffusion-limited'' or ``spatial diffusion-limited'' may be an imprecise means to distinguish the pre- and post-turnover regimes~\cite{PollakJCP1989}.}, the overall reaction rate increases with increasing $\eta_s$ as can be seen in 
Fig.~\ref{fig1}c. Finally, further increases in  $\eta_s$ drives the system into the spatial diffusion-limited regime where the reaction rate decreases with increasing solvent friction $\eta_s$. In classical rate theory this transition from energy diffusion-limited regime to the spatial diffusion-limited regime is referred to as the Kramers turnover~\cite{PeterRMP1990}. 

\begin{figure*}
\centering
\includegraphics[width=0.99\linewidth]{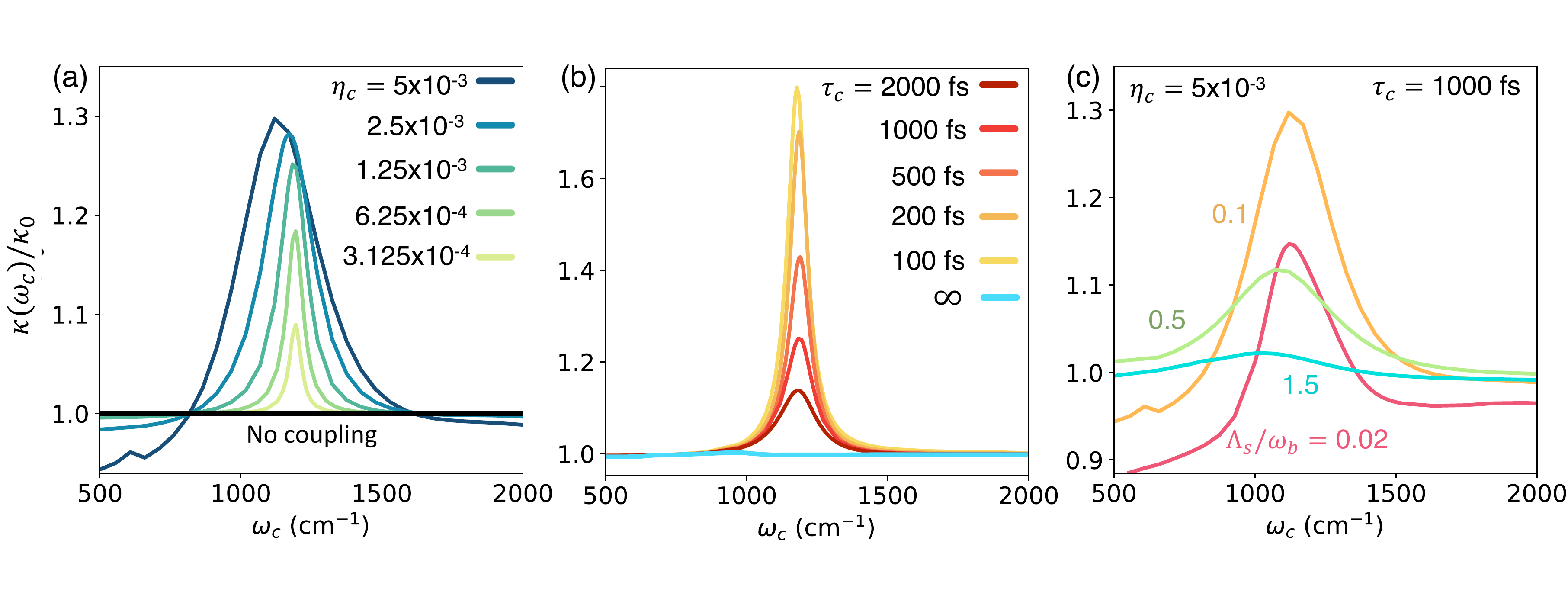}
\caption{\footnotesize \textbf{Effect of various parameters on cavity modified chemical reactivity.} (a) Effect of light-matter coupling $\eta_c$ on chemical reactivity with cavity lifetime $\tau_c = 1000$ fs and solvent friction $\eta_s = 0.1\omega_b$.  (b) Effect of cavity lifetime  $\tau_c$ at $\eta_c = 0.00125$ and $\eta_s = 0.1\omega_b$. Note the negligible cavity modifications in the absence of cavity loss $\tau_c = \infty$.  (c) Effect of solvent friction on chemical reactivity with cavity lifetime $\tau_c = 1000$ fs and light-matter coupling $\eta_c = 0.00125$. }
\label{fig3}
\end{figure*}
For a molecular system embedded in a structured solvent outside of the cavity ($\eta_c = 0$ in $\hat{H}$), the chemical rate as a function of the molecule-spectator mode coupling $C_Q$ is presented in Fig.~\ref{fig1}d. The corresponding molecular system is also schematically illustrated in Fig.~\ref{fig1}f. The effective bath spectral density~\cite{GargJCP1985,HughesJCP2009} $J_{S}(\omega) = \frac{\pi}{2} \sum_{j} \frac{\Tilde{C}_j^2}{\Tilde{\omega}_j} \delta(\omega -\Tilde{\omega}_j)$, for the effective (and equivalent) solvent Hamiltonian $\hat{H}_\mathrm{solv} \equiv \sum_j \frac{{\Tilde{p}}_{j}^2}{2} + \frac{1}{2} {\Tilde{\omega}}^2_{j} ({\Tilde{x}}_j + {\Tilde{C}_{j} {\Tilde{Q}}}/{{\Tilde{\omega}}^{2}_j})^2$, is presented in Fig.~\ref{fig1}b (red solid line). In comparison to the unstructured environment, the spectral density $J_{S}(\omega)$  for the structured environment shows a sharp spike at 1200 cm$^{-1}$, which originates from the spectator mode $Q$ with frequency $\omega_Q = 1200$ cm$^{-1}$, and the width of this peak originates from the secondary solvent spectral density $J_u(\omega)$. This is characteristic of complex molecular environment~\cite{JavierPRL2018,ChristenssonVS2010,ZuehlsdorffJPCB2020} where the spectral density contains numerous spikes. It is worth noting that the reaction coordinate is strongly coupled to other orthogonal vibrations~\cite{ClimentPCCP2020} in molecular systems considered in typical experimental work~\cite{ThomasS2019}. 

Here we will consider $\eta_s = 0.1\omega_b$, so that the system is in the weak-coupling  (pre-turnover) regime for $C_Q = 0$.  Similar to Fig.~\ref{fig1}c, in Fig.~\ref{fig1}d the reaction rate initially increases as $C_Q$ increases, and undergoes a turnover, at $C_Q\approx 3\times10^{-6}$ a.u., following which the rate decreases with increasing $C_Q$, akin to the Kramers turnover.  


{\bf  Chemical Kinetics Inside a Cavity.} 
When the molecular system is placed inside an optical cavity (schematically illustrated in Fig.~\ref{fig2}a), light-matter coupling between molecular vibrational states leads to the formation of vibrational polaritons. This is shown in Fig.~\ref{fig2}b-c. Here, we consider first a molecular system embedded in an unstructured environment (as depicted in Fig.~\ref{fig1}e) with the reaction coordinate directly coupled to the cavity radiation mode (with $\hat{\mu} = \hat{R}$). 

In the simplified Jaynes–Cummings model~\cite{JaynesPI1963}, excited vibrational states (of the ground electronic state $|G\rangle$) with 0 photons in the cavity $|v_2\rangle\otimes|G,0\rangle \equiv |v_2\rangle \otimes|0\rangle$ hybridize with the ground vibrational states with 1 photon in the cavity, $|v_L\rangle\otimes|1\rangle$, through the coupling $\langle v_2, 0|\hat{H}_\mathrm{cav}|v_L,1\rangle = \eta_\mathrm{c} \omega_c \langle v_L|\hat{R}|v_2\rangle $. This results in   an avoided crossing of polaritonic eigenenergies when $\omega_c 
\approx \omega_0 = E_{2} - \bar{E}_{0}\approx  1140 $ cm$^{-1}$ in Fig.~\ref{fig2}c, which results in a Rabi-splitting ($\sim$ 30 cm$^{-1}$) observed in absorption spectra (with $\tau_c = 100$ fs), as shown in Fig.~\ref{fig2}d (violet solid line). 

Using our HEOM approach, we have simulated the chemical kinetics of this molecule-cavity hybrid system. We have found that the chemical rate profile as a function of photon frequency $\omega_c$ is sharply peaked around 1200 cm$^{-1}$, when the cavity frequency is in resonance with the molecular vibrational transitions. Remarkably, the chemical rate profile (blue solid line in Fig.~\ref{fig2}d) has a similar lineshape compared to the absorption spectra (violet shaded area in Fig.~\ref{fig2}d). This bares a striking resemblence to recent experimental observations~\cite{ThomasACID2016,ThomasS2019,LatherACID2019,VergauweACID2019,AnoopNp2020, LatherCS2022, SimpkinsJPCC2021, NagarajanJACS2021}. We must emphasize, that current experiments operate in collective regime, while the present theoretical calculations pertain to a single-molecule single-cavity setup. Notably, however, the light-matter couplings used in this work are much smaller than  what has been used in recent theoretical work in the single molecule case~\cite{LiNC2021, LiJPCL2021, GalegoPRX2019, SunJPCL2022, PhilbinJPCC2022, ChristianArxiv2022, FabijanJACS2022}. For example, a quantization volume of  0.19 nm$^3$ was used in  Ref.~\citenum{FabijanJACS2022} while we use $V \sim $10-288 nm$^3$ here~\footnote{To make a fair comparison, here we have estimated the quantization volume $V$  by considering a proton transfer reaction, such that $\hat{\mu}  = \frac{\hat{R}}{\sqrt{m_p}}$  (where $m_p = 1836$ a.u.). Thus, the  $\eta_c  \rightarrow \sqrt{m_p}\eta_c $  is scaled up when computing the quantization volume $V$. That is, $V$ is estimated using the relation $\sqrt{m_p } \eta_c \omega_c= \sqrt{\hbar\omega_c / 2\epsilon_0 V}$.  }. This difference is significant, as it implies that the effects we observe would be manifest in real single molecule experiment and would likely be enhanced in the collective case. 

The photon frequency-dependent rate profile obtained with exact quantum dynamics  is much sharper than the rate profile obtained using direct classical simulations (using generalized Langevin equation, see Supplementary Note 3), as shown in Fig.~\ref{fig2}e. Additionally, the classical results show a significantly smaller peak enhancement that occurs at a lower frequency ($\sim$ 1000 cm$^{-1}$) than in our exact quantum calculation.  These results underscore the importance of {\it quantum} dynamical interplay of photonic and molecular degrees of freedom in modifying chemical reactivity.

In Fig.~\ref{fig3}, we analyze the role of light-matter coupling strength, cavity loss and molecule-solvent coupling strength to gain mechanistic insights into cavity modified chemical kinetics. Fig.~\ref{fig3}a presents the normalized reaction rate constant $\kappa(\omega_c)/\kappa_0$ (with $\kappa_0$ the rate constant outside cavity) at various light-matter coupling strengths $\eta_c$. Increasing the light-matter coupling increases both the width as well as the height of the rate profile. This is expected, as higher coupling increases the effective environmental friction to the molecular system in the weak coupling regime. At the same time, higher coupling also allows the cavity and vibrational excitation to hybridize at larger detuning, resulting in a wider rate profile.   

Here, we also observe an off-resonant suppression of the chemical rate at higher light-matter couplings. This effect can be understood by applying a polaron transformation leading to a rescaling of $\Delta \rightarrow \Delta \cdot \Big\langle \exp[\pm {i  \eta_c \sqrt{2/\omega_c} \hat{p}_c \Delta \mu }]\Big\rangle$ due to the difference in permanent dipoles ($\Delta \mu  = \langle v_L|R |v_L\rangle - \langle v_R|R |v_R\rangle$) between the right and left wells, resulting in a matter state-dependent displacement of the cavity mode $q_c$~\cite{Mandal2020JPCL}. Similar effects have been investigated in the context of cavity modified photo-dissociation~\cite{Mandal2020JPCL} or cavity mediated non-adiabatic electron transfer reactions~\cite{SemenovJCP2019, ChowdhuryJCP2021}.

Fig.~\ref{fig3}b presents the effects of cavity loss on the cavity modified chemical kinetics. We observe that an increase in cavity loss leads to a significant increase in the reaction rate at resonance, but leads to only minor changes off-resonance. The effect can be accounted for by considering the energy transfer processes occurring between the molecule-cavity subsystem (containing only $R$ and $q_c$ and described by $\hat{H}_\mathrm{mol} + \hat{H}_\mathrm{cav}$) when coupled to a dissipative bath composed of the solvent degrees of freedom $\{X_j\}$ and how the presence of far-field cavity modes $\{\mathcal{Q}_{k}\}$ modifies these processes. In the absence of cavity loss, thermalization of the cavity mode can only occur through energy transfer between the cavity mode and the molecular bath that are mediated by the reaction coordinate.  As a consequence, the cavity mode does not provide an efficient mechanism for energy loss, as excess energy transferred to the cavity mode during reaction will necessarily transfer back to the reaction coordinate as the system approaches equilibrium.  The inclusion of far-field modes that couple to the cavity mode in $\hat{H}_\mathrm{loss}$ provide an additional pathway for thermalization of the cavity mode that does not require energy transfer through the reaction coordinate. This allows for the cavity mode to act as an additional source of environmental friction for the reaction coordinate, which leads to the enhancement of the rate constant kinetics in the weak-coupling regime. 



Consequently, we find that cavity modification of the reaction rate is negligible in the absence of cavity loss ($\tau_c \rightarrow \infty$, solid blue line) and shows no resonance structure (see further details in Supplementary Note 4). Therefore, cavity loss plays a significant role in modifying chemical reactivity in this regime. 

\begin{figure*}
\centering
\includegraphics[width=1.0\linewidth]{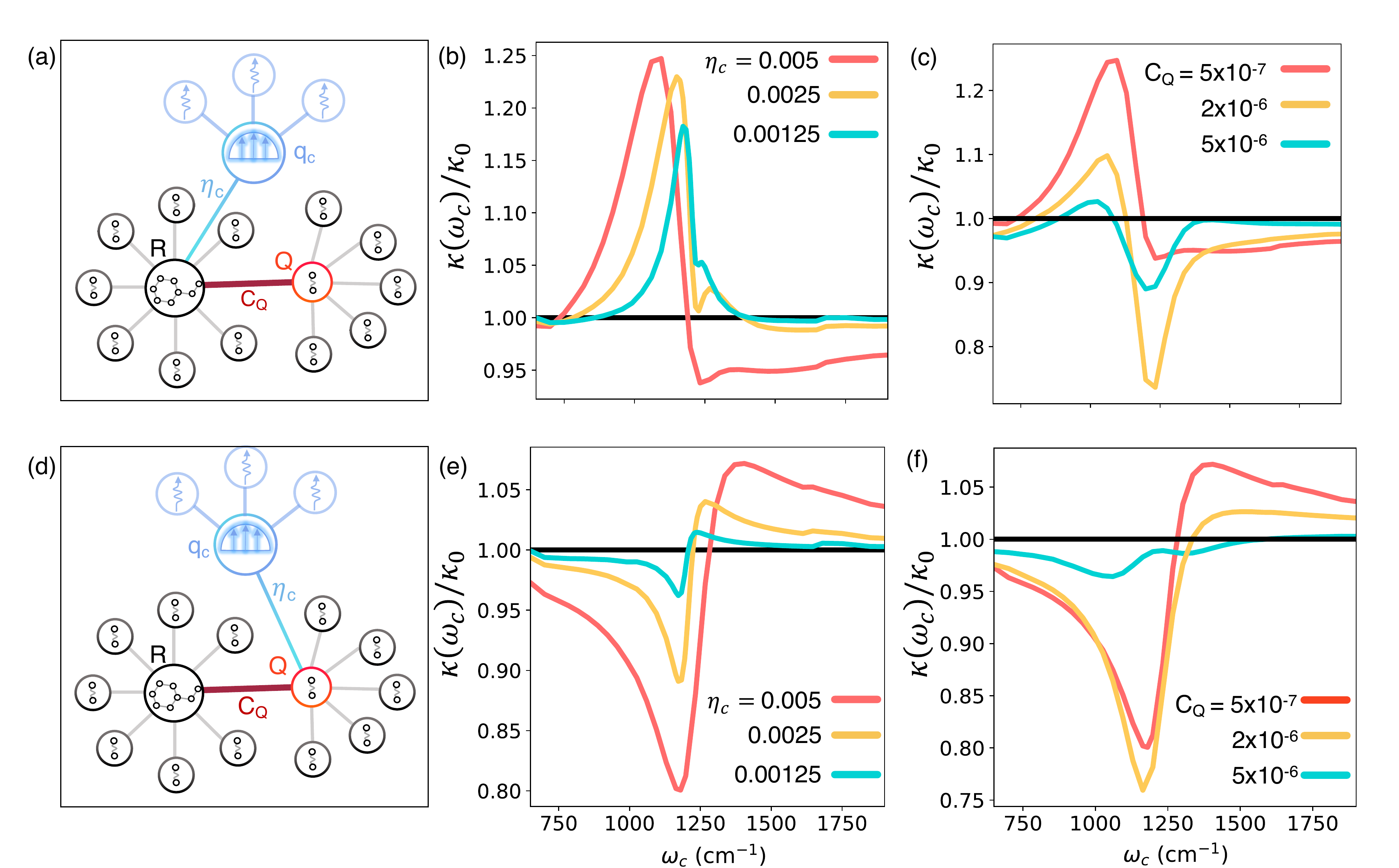}
\caption{\footnotesize \textbf{Cavity modification of chemical reactivity in complex molecular systems.} (\textbf{a}) Schematic illustration of molecular system embedded in a structured environment and coupled to a lossy cavity. Cavity modification of chemical reactivity at (\textbf{b}) various light-matter coupling values $\eta_c$ with fixed $C_Q = 5\times 10^{-7}$ a.u. and at various values of (\textbf{c}) $C_Q$ with fixed  $\eta_c = 0.005$ a.u. (\textbf{d}) Similar to (\textbf{a}) but with the cavity coupling to the spectator mode $Q$. (\textbf{e}) and (\textbf{f}) are similar to \textbf{b} and \textbf{c} but for model system illustrated in (\textbf{d}). The cavity lifetime is set to $\tau_c = 1000$ fs.}
\label{fig4}
\end{figure*}
In Fig.~\ref{fig3}c we investigate the role of the molecule-solvent interaction strength on the cavity-modified reaction rate at  $\eta_c = 0.005$ and $\tau_c = 1000$ fs. For very weak coupling, $\eta_s = 0.02\omega_b$ a.u. (red solid line), we observe off-resonant suppression as well as resonant enhancement of the chemical reaction rate. Physically, this occurs due to the fact that for low $\eta_s$ both tunneling as well as thermal activation (such as in the energy diffusion-limited regime) reaction pathways play important roles in the chemical reaction. The cavity suppresses the tunneling by renormalizing $\Delta$ as discussed above, while at the same time resonantly enhancing the thermal activation pathway by effectively increasing the environmental friction. The combination of these two competing effects leads to both off-resonant suppression and resonant enhancement of the rate at $\eta_s = 0.02\omega_b$. For $\eta_s = 0.1\omega_b$, as tunneling plays a less significant role, off-resonant suppression becomes less prominent. Thus at $\eta_s = 0.1\omega_b$, the cavity modified rate constant primarily shows resonant enhancement. Further increase in $\eta_s$ decreases the extent of cavity modification as higher $\eta_s$ places the system closer to the turnover region. Thus, for  $\eta_s = 1.5\omega_b$ (near the Kramers turnover, see Fig.~\ref{fig1}c), the cavity modification becomes negligible. Due to the computational expense of our HEOM calculations, simulation of the cavity-modified reaction rate is outside our reach for $\eta_s > 1.5\omega_b$. However, we speculate that cavity mediated suppression could be observed at higher $\eta_s$.  This will be explored using more efficient but approximate quantum dynamics approaches in future work. 

So far we have investigated a molecular system embedded in an unstructured environment. In  Fig.~\ref{fig4} we consider the molecular system embedded in a structured environment as depicted in Fig.~\ref{fig1}f,  characteristic of real molecular systems. Here we consider two scenarios, one where the cavity radiation mode is directly coupled to the reaction coordinate $R$, and one where it directly couples to the spectator mode $Q$, as schematically illustrated in Fig.~\ref{fig4}a and Fig.~\ref{fig4}d, respectively. 

In  Fig.~\ref{fig4}b we present the effect of the cavity coupling, at a fixed value of the of coupling to the spectator mode $C_Q = 5\times 10^{-7}$ a.u., molecule-solvent coupling $\eta_s = 0.1\omega_b$ and a cavity lifetime of $\tau_c = 1000$ fs, on the frequency-dependent chemical reaction rate for the cavity setup illustrated in Fig.~\ref{fig4}a. Here, several competing effects play a role in modifying chemical reactivity. First, directly coupling $R$ to the cavity (which is coupled to a set of dissipative modes $\{\mathcal{Q}_k\}$) increases the effective environmental friction,  leading to a resonant enhancement of the reaction rate which peaks around 1140 cm$^{-1}$ as discussed previously and shown in Fig.~\ref{fig2}-\ref{fig3}. Secondly, coupling to the cavity splits the molecular vibrational levels (creating polariton states), pushing them further away from the sharp peak of the ($\omega_{Q} = 1200$ cm$^{-1}$)  structured spectral density (red solid line in Fig.~\ref{fig1}b), leading to a suppressed interaction between $R$ and $Q$. In other words, coupling to a cavity can drive the reaction coordinate away from resonance with the relevant solvent/spectator modes, thereby {\it reducing the effective environmental friction}. 

Further, note that $\omega_0 \approx  1140 $ cm$^{-1}$ is slightly lower than  $\omega_{Q} = 1200$ cm$^{-1}$. As a result, when $\omega_c < \omega_0$, the chemical reaction rate is enhanced due to the enhanced interaction between $R$ and $Q$ as $|v_{2},0\rangle$ is pushed up energetically from below. On the other hand, for $\omega_c > \omega_0$ the molecular vibrational states (i.e. the lower polariton states) are pushed down, away from $\omega_{Q}$, reducing the interaction between $R$ and $Q$. In the Supplementary Information we consider the opposite scenario where $\omega_{Q} < \omega_0$, such that the shape of $\kappa(\omega_c)$ is reversed, i.e. suppression at lower $\omega_c$ and enhancement at higher  $\omega_c$,  supporting this picture.

At higher light-matter coupling strengths, due to the larger Rabi-splitting, the suppression of interactions between ${Q}$ and $R$ becomes more important. Here, the photon frequency-dependent rate constant at relatively high light-matter coupling, $\eta_c = 0.005$ (red solid line in Fig.~\ref{fig4}b), shows suppression at higher photon frequencies but enhancement at lower photon frequencies. However, at relatively weaker light-matter interactions enhancement of the chemical rate, primarily originating from the effective increase in environmental friction, is observed. This can be seen for $\eta_c = 0.0025$ and $\eta_c = 0.00125$ a.u. depicted by the yellow and green line in Fig.~\ref{fig4}b, respectively. The chemical rate profile at  $\eta_c = 0.00125$ (green solid line in Fig.~\ref{fig4}b) is similar to that of the unstructured case (when $C_Q = 0.0$) shown in Fig.~\ref{fig3}b with the same cavity lifetime $\tau_c =1000$ fs, which is in harmony with the reasoning presented above.

Fig.~\ref{fig4}c illustrates the cavity modification of the chemical rate at various $C_Q$ and at constant $\eta_c = 0.005$ and $\eta_s = 0.1\omega_b$. The two competing effects and their relative importance  depends on the  value of $C_Q$. At higher $C_Q =2\times 10^{-6}$ a.u. (yellow solid line in Fig.~\ref{fig4}c), the suppressed interaction between ${Q}$ and $R$ plays a more important role, which results in a sharp  suppression of the reaction rate around $1200$ cm$^{-1}$. Interestingly, at even higher $C_Q = 5\times 10^{-6}$ a.u., the extent of cavity suppression of chemical reactivity becomes smaller. Note that with $C_Q = 5\times 10^{-6}$ a.u., the system is just past the Kramers turnover point (see Fig.~\ref{fig1}d). As a result, the decrease in the effective environmental friction (due to the light-matter Rabi-splitting) does not provide a significant modification of the chemical reaction rate.


Finally in Fig.~\ref{fig4}d-f, the cavity modification of chemical reactivity is investigated when coupling the cavity radiation mode to the spectator mode $Q$, namely, using $\hat{\mu} = \hat{Q}$ in $\hat{H}_\mathrm{cav}$. The cavity-molecule setup is schematically illustrated in Fig.~\ref{fig4}d. Coupling the spectator mode to the cavity leads to the creation of polaritonic normal modes~\cite{JorgeJCP2020}, which results in the splitting of the peak in the structured spectral density $J_S(\omega)$ in Fig.~\ref{fig1}b. Due to this effect, the peak of the spectral density can be shifted away from the $|v_L\rangle \rightarrow |v_2\rangle$ transition, leading to a suppressed interaction between the molecule and its environment (which now also include the cavity mode).  

In Fig.~\ref{fig4}e the effect of increasing the light-matter coupling $\eta_c$ between $Q$ and $q_c$ is investigated. Overall, increasing the light-matter coupling result in a larger splitting of the spectral density, leading to larger amount of suppression, with a dip occurring around $\omega_Q = 1200$ cm$^{-1}$.  As mentioned previously, since $\omega_Q$  is slightly higher than the $\omega_0 = E_{2} - \Bar{E}_{0} \approx 1140$ cm$^{-1}$ the peak of the spectral density $\omega_Q$ is pushed closer to the molecular energy gap $\omega_0$  for $\omega_c > \omega_Q$. This results in an increased interaction between the molecule and its environment at higher photon frequencies ($>1250$ cm$^{-1}$), leading to an enhancement of chemical reactivity in Fig.~\ref{fig4}e in the pre-turnover, weak-coupling regime.

In Fig.~\ref{fig4}f we explore the effect of increasing $C_Q$ on cavity modified chemical reactivity. Similar to Fig.~\ref{fig4}c, we observe that an initial increase in $C_Q$ (to $2\times 10^{-6}$ a.u., yellow solid line) leads to further modification (i.e. suppression) of chemical kinetics. This behavior can be understood by looking at how the reaction rate is modified as function of $C_Q$, as presented in Fig.~\ref{fig1}d. For very low $C_Q \rightarrow 0$, the slope of the rate $\kappa$ is smaller than that of the slope at relatively higher values $C_Q = 2\times 10^{-6}$ a.u. This implies that changes in environmental friction for $C_Q = 5\times 10^{-7}$ a.u. lead to small changes in the chemical reaction rate than they do for $C_Q = 2\times 10^{-6}$ a.u. This effect explains why chemical rate modifications are relatively milder for $C_Q = 0.5\times 10^{-5}$ a.u. which is close to the turnover region  (see Fig.~\ref{fig1}d). Importantly, the resonant structure of the rate profile appears due to the modification of the environmental friction which originates from the  Rabi splitting caused by quantum light-matter interactions. 

Overall, the  results presented in Fig.~\ref{fig4} demonstrate that it is possible to observe both resonant enhancement or resonant suppression depending upon how the cavity couples to the matter subsystem. The cavity can either enhance (such as in \ref{fig4}a-c) or suppress (such as in \ref{fig4}d-f) the effective environmental friction felt by the molecule. These effects originate from quantum light-matter hybridization, and as a consequence the chemical rate as a function of photon frequency exhibit sharp resonances.  

\begin{figure}
\centering
\includegraphics[width=1.0\linewidth]{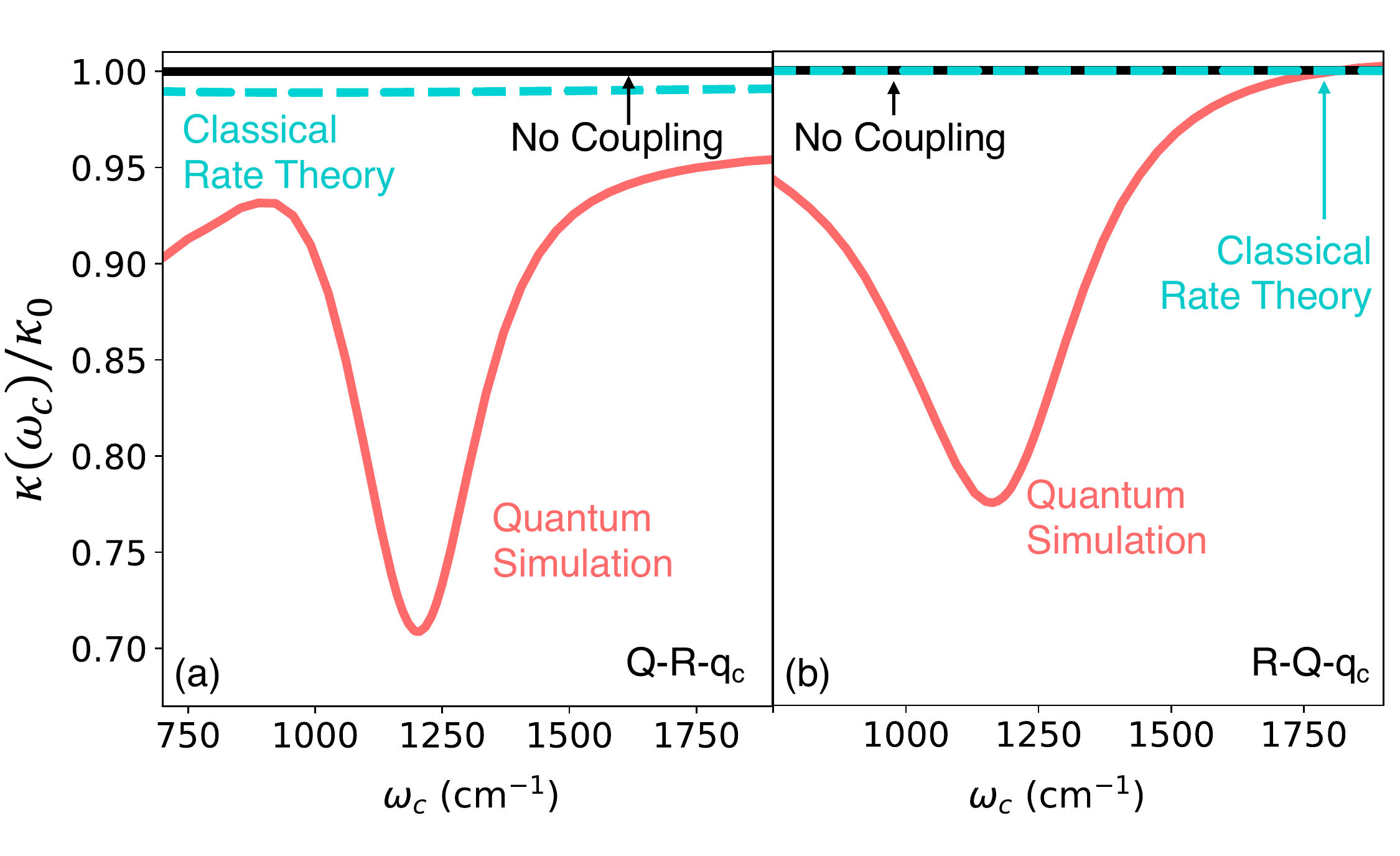}
\caption{\footnotesize \textbf{Resonant cavity suppression of chemical kinetics under strong (post-turnover) solvent-molecule interactions.} Chemical rate constant $\kappa$ as a function of $\omega_\mathrm{c}$ for a molecular system embedded in a structured environment (same as Fig.~\ref{fig4}) when the cavity mode $q_\mathrm{c}$ is coupled to ({a}) the reaction coordinate $R$ (illustrated in Fig.~\ref{fig4}a) or ({b}) to the spectator mode $Q$ (illustrated in Fig.~\ref{fig4}d) computed using exact quantum dynamics (red solid line) and the classical (Grote-Hynes theory~\cite{GroteJCP1980, PeterRMP1990}) rate theory (dashed cyan line) compared with the chemical rate constant in the absence of the cavity (black solid line). The molecule-spectator mode coupling $C_Q = 5\times 10^{-6}$ a.u. places the system in the post-turnover regime (using an unstructured solvent friction $\eta_s = 0.1 \omega_b$). The cavity lifetime is set to $\tau_c = 1000$ fs and the light-matter coupling is set to $\eta_c = 0.01$ a.u.} 
\label{fig5}
\end{figure}

Finally, in Fig.~\ref{fig5} we demonstrate that sharp resonant suppression of ground state chemical kinetics is observed even when molecule-solvent interactions are in the strong coupling (post-turnover) regime. This is important, since many systems are experimentally expected to be found in this strong-coupling (post-turnover) regime. Here, the cavity mode $q_c$ is coupled to $R$ in Fig.~\ref{fig5}a or to $Q$ in Fig.~\ref{fig5}b as schematically illustrated in Fig.~\ref{fig4}a and d respectively.  Importantly, here we set $C_Q = 5\times 10^{-6}$ a.u. (and $\eta_s = 0.01\omega_b$) that places the environmental friction in the post-turnover regime. Despite the strong coupling (post-turnover), a significant suppression of chemical reactivity, in the range of $25-30\%$, is achieved in both scenarios by using a slightly higher light-matter coupling $\eta_c = 0.01$ a.u. (corresponding to a quantization volume of $V\sim 10$ nm$^{3}$).  In both cases, our exact quantum simulations show a sharp resonant suppression, just as observed in experiments~\cite{ThomasS2019,ThomasACID2016, NagarajanJACS2021} and in contrast to all previous theoretical studies that show a weak and broad cavity suppression profile~\cite{GalegoPRX2019, LiNC2021, LiJPCL2021, ChristianArxiv2022, YangJPCL2021, MandalJCP2022, FischerJCP2022, PhilbinJPCC2022, SunJPCL2022}. Here we compare the rate constant computed from the quantum simulations (red solid line) with Grote-Hynes classical rate theory~\cite{PeterRMP1990, GroteJCP1980, LiNC2021} which predicts minute and broad suppression of chemical reactivity. Note that the quantum results presented in Fig.~\ref{fig5}a show a cavity frequency-independent suppression in addition to the resonant cavity frequency dependent suppression. This cavity frequency independent suppression is due to the renormalization (rescaling $\Delta$) of the tunneling parameter when coupling $q_c$ directly to the reaction coordinate $R$, as explained in Fig.~\ref{fig3}.


{\bf Summary and Conclusion. } In this work we present an in-depth study of cavity modification of chemical reactivity in a model that contains most of the features needed to realistically describe how light-matter coupling may alter chemical reaction rates.  Crucially, our work treats all degrees of freedom quantum mechanically, and does so in a numerically exact manner.  This enables a detailed understanding of the factors that enhance or reduce reaction rates and the explicit role played by Rabi splitting and resonances on these effects.  Our work highlights the fact that {\em the quantitative description of the modification of the rate of a chemical reaction inside an optical cavity requires a full treatment of the quantum nature of the cavity radiation mode and molecular vibrations.}

Remarkably, we find that the features found in our simulations are similar to those seen in experiments carried out at finite molecular concentrations.  We show that alteration of reactivity is subtle, and depends sensitively on factors such as which modes couple to the cavity and the rate of cavity loss.  Many aspects of cavity-modified reactivity that we expose in this work have not been previously captured by theoretical studies.  In particular, we find that most enhancement or suppression of chemical reactions occurs near resonance with molecular vibrational frequencies, and that these features are sharp (e.g., on the order of a few hundred wave numbers in width at most).  For example, whereas past classical and semiclassical calculations could produce only a very weak and very broad suppression of the rate in the regime past the Kramers turnover, we find that sizable and sharp suppression may occur in this regime once an exact, fully quantum approach is used.  In fact the magnitude of the effects we see are surprisingly large given that we employ realistically small couplings and realistic levels of cavity loss.

One obvious important aspect missing from our model is collective effects induced by the cavity and by direct coupling between molecules.  It is natural to assume that collective effects will increase the magnitude of the features presented here without qualitatively altering the behavior. However explicit many-molecule quantum mechanically calculations are needed to confirm this expectation.  An exact quantum mechanical treatment of such a situation is likely prohibitively expensive for realistic models. On the other hand, the results presented here may be used as benchmarks for approximate quantum mechanical approaches aimed at treating such collective effects.  We plan to carry out such studies in future work.

\section{Methods}
{\bf Hierarchical Equations of Motion.}
All Quantum mechanical simulations were performed using the Hierarchical Equations of Motion (HEOM) approach.  This well-established open-quantum system dynamics method provides an exact description of the dynamics of a quantum system that is linearly coupled to a set of $N$ harmonic baths \cite{doi:10.1143/JPSJ.58.101,doi:10.1143/JPSJ.74.3131,     doi:10.1063/5.0011599}.  The Hamiltonian for such a system may be written as
\begin{equation}
    \hat{H} = \hat{H}_{S} + \sum_{i=1}^N \hat{H}_{B,i} + \sum_{i=1}^N \hat{S}_i \hat{B}_i, \label{eq:heom_hamiltonian}
\end{equation}
where $\hat{H}_S$ and $\hat{H}_{i,B}=\sum_{\alpha}\omega_{i,\alpha} \hat{a}_{i,\alpha}^\dagger\hat{a}_{i,\alpha}$ are the bare system and $i$th harmonic bath Hamiltonians, respectively, $\hat{S}_i$ is a system operator, and $\hat{B}_i=\sum_{\alpha} g_{i,\alpha} (\hat{a}_{i,\alpha}^\dagger+\hat{a}_{i,\alpha})$ is the $i$th bath coupling operator.  The frequencies, $\omega_{i,\alpha}$, and coupling constants, $g_{i,\alpha} = c_{i\alpha}/\sqrt{2\omega_{i\alpha}}$ for bath $i$ are fully specified by the bath spectral density $J_i(\omega) = \pi \sum_\alpha g_{i,\alpha}^2 \delta(\omega-\omega_{i,\alpha})$, which uniquely captures the influence of the bath on the system at a given temperature.

The HEOM approach makes use of the Feynman-Vernon influence functional to obtain the exact quantum dynamics of the system in terms of an infinite set of auxiliary density operators (ADOs), $\hat{\rho}_{\boldsymbol{n}}(t)$ \cite{doi:10.1143/JPSJ.58.101,doi:10.1063/5.0011599}.  These ADOs are indexed by sets of integers of length $K_i$ for each bath, that is $\boldsymbol{n} = (n_{0,0}, n_{0, 1}, \dots, n_{0, K_0}, n_{1, 0}, \dots, n_{N, K_N})$.  This infinite set of ADOs contains the system reduced density operator as the element $\hat{\rho}_{\boldsymbol{0}}(t) = \mathrm{Tr}_B \left[\hat{\rho}(t)\right]$, where $\mathrm{Tr}_B$ denotes the partial trace over all bath degrees of freedom and $\hat{\rho}(t)$ is the density operator for the full system and bath.  All other ADOs account for correlations between the system and bath degrees of freedom that arise from the system bath coupling terms in Hamiltonian Eq. \ref{eq:heom_hamiltonian}.

The ADOs evolve according to the infinite set of coupled ordinary differential equations \cite{doi:10.1063/1.3077918}
\begin{equation}
\begin{split}\label{eq:HEOM}
    &\frac{\mathrm{d}}{\mathrm{d}t} \hat{\rho}_{\boldsymbol{n}}(t) = -i\left[\hat{H}_S, \hat{\rho}_{\boldsymbol{n}}(t)\right] - \sum_{i=1}^N \sum_{k=1}^{K_j} n_{i,k} \nu_{i,k} \hat{\rho}_{\boldsymbol{n}}(t) \\
&\hspace{-6pt}-\!i\sum_{i=1}^N \sum_{k=1}^{K_j}\! \left(\!\sqrt{n_{i,k}\!+\!1} \mathcal{B}_{i,k}^-\hat{\rho}_{\boldsymbol{n}_{i,k}^+}\!(t)\!+\! \sqrt{n_{i,k}}\mathcal{B}_{i,k}^+\hat{\rho}_{\boldsymbol{n}_{i,k}^-}\!(t)\right), 
\end{split}
\end{equation}
where $\boldsymbol{n}_{i,k}^\pm = (n_{0,0}, \dots, n_{i, k}\pm 1, \dots, n_{N, K_N})$, and here we have used a rescaled form for the ADOs.  In Eq. \ref{eq:HEOM} the Liouville space system operators, $\mathcal{B}_{i,k}^{\pm}$, are defined by
\begin{align}
    \mathcal{B}_{i,k}^{-}\hat{O} &=  \sqrt{|c_{i,k}|} \left[\hat{B}_i, \hat{O}\right] \\
    \mathcal{B}_{i,k}^{+}\hat{O} &= \left( \frac{c_{i,k}}{\sqrt{|c_{i,k}|}}\hat{B}_i \hat{O} - \frac{\overline{c}^*_{i,k}}{\sqrt{|c_{i,k}|}} \hat{O} \hat{B}_i \right)\\
\end{align}
where $\nu_{i,k}$, $c_{i,k}$ and $\overline{c}_{i,k}$ are obtained from a decomposition of the bath correlation function,
\begin{equation}
C_i(t) = \frac{1}{Z_b}\mathrm{Tr}\left[\hat{B}_i (t) \hat{B}_i e^{-\beta \hat{H}_{i,B}}\right],
\end{equation}
and its complex conjugate into a sum of exponentials \cite{doi:10.1063/5.0011599},
\begin{align}
    C_i(t) &= \sum_{k=1}^{K_i} c_{i,k} e^{-\nu_{i,k} t} \\
    C_i^*(t) &= \sum_{k=1}^{K_i} \overline{c}^*_{i,k} e^{-\nu_{i,k} t}.
\end{align}
For an exact representation of the bath correlation function it will typically be necessary to include an infinite number of terms in this expansion.  Here we truncate to a finite number of terms $K_i$ for each bath by using a ($[K_i-1]/[K_i]$) P\'ade approximant of the Bose function \cite{doi:10.1063/1.3484491,doi:10.1063/1.3602466}.  

In order to obtain a practical simulation method it is typical to truncate this infinite series of ADOs.  For the models considered in the main text, and in particular those approaching the Kramer's turnover regime, we were unable to converge the dynamics with respect to the size of the hierarchy using the commonly used strategies of truncating the hierarchy of ADOs at a fixed total excitation number \cite{doi:10.1143/JPSJ.74.3131}, $L$,  (including all ADOs satisfying the conditions $\sum_{i=1}^N\sum_{k=1}^{K_i} n_{i,k} \leq L)$ or at a fixed total decay rate\cite{doi:10.1063/1.4997433}, $\nu_M$, (including all ADOs satisfying $\sum_{i=1}^N\sum_{k=1}^{K_i} \nu_{i,k} n_{i,k} \leq \nu_M$).  We found that converged results could be obtained using a modified truncation strategy in which we include all ADOs satisfying the condition
\begin{equation}
    \sum_{i=1}^N\sum_{k=1}^{K_i} \mathrm{max}\left(\nu_{i,k} n_{i,k}, \Gamma \nu_{\mathrm{min}}\right) \leq L \nu_{\mathrm{min}},
\end{equation}
where $\nu_{\mathrm{min}} = \mathrm{real}(|\nu_{i,k}|)$. 
This approach combines aspects of the two more commonly used approaches-it includes all ADOs:
\begin{itemize}
    \item with decay rate less than $\nu_M = L \nu_{min}$\vspace{-6pt}
    \item and with up to $L/\Gamma$ excitations regardless of their decay rate
\end{itemize}

{\bf Evaluation of the Forward Reaction Rate.}  In order to evaluate the forward reaction rate, we have assumed that the system is initially in the reactant region with an initial density operator $\hat{\rho}(0) = \hat{\rho}_R$.  The time-dependent reactant and product populations may be written as
\begin{equation}
\begin{aligned}
    P_R(t) &= \mathrm{Tr}\left[(1-\hat{h})\hat{\rho}(t)\right]\\
    P_P(t) &= 1 - P_R(t),
\end{aligned}
\end{equation}
where $\hat{h}$ is the side operator that projects onto the reactant states, and is only a function of the reaction coordinate position operator, $\hat{R}$, in our work.  Provided first-order kinetics provides a valid description of the reaction process, then in the long-time limit the reactant and product populations will evolve according to the kinetic equations \cite{doi:10.1063/1.5116800, doi:10.1063/5.0098545, QiangJCP2011}
\begin{equation}
\begin{aligned}
\dot{P}_R(t) &= -\kappa P_R(t) + \kappa' P_P(t) \\
    \dot{P}_P(t) &= \kappa P_R(t) - \kappa' P_P(t),
\end{aligned}
\end{equation}
where $\kappa$ and $\kappa'$ are the forward and backward rate constants, respectively (and are related by $\kappa \langle P_R\rangle = \kappa' \langle P_P \rangle$, where $\langle P_R\rangle$ and $\langle P_P \rangle$ are the equilibrium reactant and product populations).  Rearranging the expression for the forward rate constant, we have \cite{doi:10.1063/1.2772265, doi:10.1063/1.5116800, QiangJCP2011}
\begin{equation}\label{Eq:RateConstant}
    \kappa = \lim_{t\rightarrow\infty}\frac{\dot{P}_P(t)}{1-P_P(t)/\langle P_P\rangle},
\end{equation}
where the limit $t\rightarrow\infty$ indicates that the kinetic description of the reaction process is only valid after some initial transient process.

We have considered two choices for the reactant density operator.  The first, an uncorrelated (between system and bath), thermal density operator \cite{QiangJCP2011}
\begin{equation}
    \hat{\rho}_R = \frac{1}{Z_R}e^{-\beta \hat{H}_S/2} (1-\hat{h})e^{-\beta \hat{H}_S/2} \otimes \frac{e^{-\beta \hat{H}_B}}{\mathrm{Tr}\left[e^{-\beta \hat{H}_B}\right]}, 
\end{equation}
where $Z_R = \mathrm{Tr}\left[e^{-\beta \hat{H}_S/2} (1-\hat{h})e^{-\beta \hat{H}_S/2} \right]$, allows for the direct application of the HEOM approach discussed above.  The second, a correlated, thermal reactant density operator \cite{doi:10.1063/1.2772265, doi:10.1063/5.0010580}
\begin{equation}
    \hat{\rho}_R = \frac{\hat{\rho}_{SS} (1-\hat{h})}{\mathrm{Tr}\left[\hat{\rho}_{SS}(1-\hat{h})\right]} ,
\end{equation}
where we have used the steady state solution of the HEOM,  $\hat{\rho}_{SS}$, as the thermal equilibrium state \cite{doi:10.1063/1.4890441}, requires the evaluation of the HEOM steady state before evaluation of the rate constant.  
The short-time transient dynamics depends on the choice of the initial reactant density operator, however, we have found that the long-time plateau value of Eq. \ref{Eq:RateConstant} is independent of the choice of initial density operator for the models considered in this work.

\section {\normalsize Data Availability}
    The data that support the plots within this paper and other findings of this study are available from the corresponding authors upon a reasonable request.
\section {\normalsize Code Availability}

The source code that support the findings of this study are available from the corresponding author upon reasonable request.

\section {\normalsize Acknowledgments}
This work was supported by NSF-1954791 (A.M. and D.R.R.) and by the Chemical Sciences, Geosciences, and  Biosciences Division of the Office of Basic Energy Sciences, Office of Science, U.S. Department of Energy (L.P.L. and D.R.R.). This work used the Extreme Science and Engineering Discovery Environment
(XSEDE), which is supported by National Science Foundation grant number ACI-1548562
(allocations: TG-CHE210085). Specifically, it used the services provided by the OSG Consortium,
which is supported by the National Science Foundation awards $\#$2030508 and $\#$1836650.
 
\section{\normalsize Author contributions}
L. P. L., A. M., and D. R. R. designed the research. L. P. L. performed the exact quantum dynamics simulations. A. M. performed classical simulation. L. P. L. and A. M. set up the model for molecule-cavity system. L. P. L., A. M., and D. R. R. wrote the manuscript.

\section{\normalsize Competing Interests}
The authors declare no competing interests.

\bibliography{bib.bib}
\bibliographystyle{naturemag}

\end{document}


\title{Supplementary Information: Quantum Dynamics of Vibrational Polariton Chemistry}

\author{Lachlan P. Lindoy$^1$, Arkajit Mandal$^1$, David R. Reichman$^1$~\footnote{drr2103@columbia.edu}}

\maketitle

{ $^1$ Department of Chemistry, Columbia University, 3000 Broadway, New York, New York, 10027,  U.S.A}

 {

  \section{Supplementary Note 1: Details of Molecular Systems}
   In the main-text we consider a one dimensional reaction coordinate coupled to a set of dissipative modes, such as ${Q}$ and $\{X_j\}$. A set of secondary dissipative modes $\{x_j\}$ and $ \{\mathcal{Q}_k\}$ are also coupled to the spectator mode $Q$ and the cavity mode $q_c$. In the main text we consider three Debye baths that are coupled: to the reaction coordinate described by the spectral density $J_U(\Omega) = 2{\Lambda}_{s}\Omega{\Gamma}/(\Omega^2+ {\Gamma}^2)$ (the solvent bath), to the spectator mode described by the spectral density $J_u(\omega) = 2\lambda_{s}\omega\gamma/(\omega^2+\gamma^2)$, and to the cavity mode described by the spectral density ${J}_{L}(\omega) =  2\lambda_{L}\omega\gamma_{L}/(\omega^2+\gamma_{L}^2)$ (the cavity loss bath).  In the main-text the reorganization energy of the reaction coordinate bath, $\Lambda_s$, was obtained by setting the solvent friction, $\eta_s$, associated with the bath $\Lambda_s = \Gamma\eta_s/2$.  The bath reorganization energies for the four friction values considered in Fig. 3c of the main text are provided below
      \begin{center}
\begin{tabular}{ |c|c|c|c|c| } 
\hline
 $\eta_s$ & $0.02\omega_b$ & $0.1\omega_b$ &  $0.5\omega_b$ &  $1.5\omega_b$  \\
\hline
   $\Lambda_s$ & $4.15\times10^{-8}$ a.u. & $2.08\times10^{-7}$ a.u.  & $1.04\times10^{-6}$ a.u. & $3.11\times10^{-6}$ a.u.\\
\hline
\end{tabular}
\end{center}
   The reorganization energy for the cavity loss bath was obtained by setting the cavity loss parameter, $\tau_c$, to $1/\tau_c = 2 J_L(\omega_{c})/(1- e^{-\beta\omega_c}   )$, and as such the reorganization energy for cavity loss depends on both the cavity loss parameter and the cavity frequency, $\omega_c$.  All other parameters, are provide below:
   \begin{center}
\begin{tabular}{ |c|c|c|c| } 
\hline
 $\lambda_s$ & $\Gamma$ & $\gamma$ & $\gamma_L$   \\
\hline
   $6.70\times10^{-7}$ a.u. & 200 cm$^{-1}$ & 1000 cm$^{-1}$ & 1000 cm$^{-1}$\\
\hline
\end{tabular}
\end{center}

  \section{Supplementary Note 2: Various Molecular Systems}
  In this section we provide additional numerical results.

\subsection{Additional Unstructured Spectral Density Results}
  \subsubsection{Low Frequency Symmetric Model Potential}
  
  \begin{figure*}[h]
\centering
\includegraphics[width=0.99\linewidth]{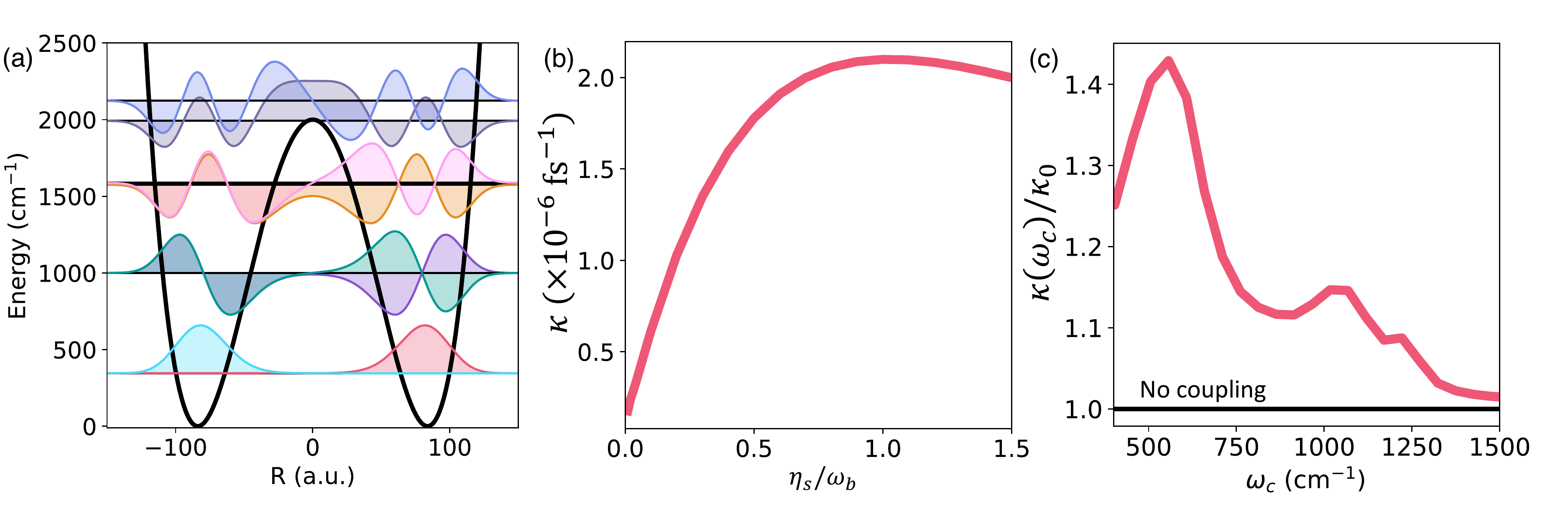}
\caption{ \textbf{Cavity modification of ground state chemical kinetics.} (\textbf{a})   Potential energy surfaces for molecular ground adiabatic state with vibrational eigenstates. (\textbf{b}) Chemical rate constant as function of solvent friction. (\textbf{c}) Cavity modified chemical rate constant as function of cavity photon frequency $\omega_c$.   }
\label{fig1}
\end{figure*}

  In Fig. \ref{fig1}, we present cavity frequency-dependent rate profiles obtained for a model with an unstructured spectral density with friction constant $\eta_s  = 0.1\omega_b$ and for a symmetric reaction coordinate potential with a lower barrier frequency $E_b = 2000$ cm$^{-1}$ and lower well frequency, $\omega_b = 500$ cm$^{-1}$  than considered in the main text.  As shown in Fig. \ref{fig1}a, this potential supports three vibrational modes in each well below the barrier.  The rate constant as a function of the bath friction is shown in Fig.~\ref{fig1}b. Here we see the characteristic Kramer's turnover, and in contrast to the results in the main text we observe no tunneling dominated regime at low friction.  The cavity frequency dependent rate constant, obtained with $\eta_c = 0.005$ a.u. and a cavity loss of $\tau_c = 1000$ cm$^{-1}$, shown in Fig. \ref{fig1}c shows a number of distinct peaks

  \subsubsection{High Frequency Symmetric Model Potential}

    In Fig. \ref{fig2} we present cavity frequency-dependent rate profiles obtained for a model with an unstructured spectral density with friction constant $\eta_s  = 0.1\omega_b$ and for a symmetric reaction coordinate potential with a higher barrier frequency $E_b = 2500$ cm$^{-1}$ and higher well frequency, $\omega_b = 1500$ cm$^{-1}$   than  considered in the main text.  As shown in Fig. \ref{fig2}a, this potential supports only a single vibrational modes in each well below the barrier, with the first excited modes being found above the barrier.  The rate constant as a function of the bath friction is shown in Fig.~\ref{fig2}b, here we see a sharp decrease in the rate constant with increasing rate constant that is characteristic of the tunneling dominated regime \cite{QiangJCP2011}.  Additionally, even at large friction constants, past the Kramer's turnover point for the other models, we do not see any turnover, instead we observe  a decrease in the rate constant with increasing friction.  For this model, no significant, resonant cavity modification of the reaction rate is observed.  As shown in Fig.~\ref{fig2}c, where we have included a cavity mode that couples to the system with $\eta_c = 0.005$ a.u. and a cavity loss of $\tau_c = 1000$ cm$^{-1}$, the presence of the cavity mode leads to an off-resonant suppression as has been discussed in the main text.

  \begin{figure*}[h]
\centering
\includegraphics[width=0.99\linewidth]{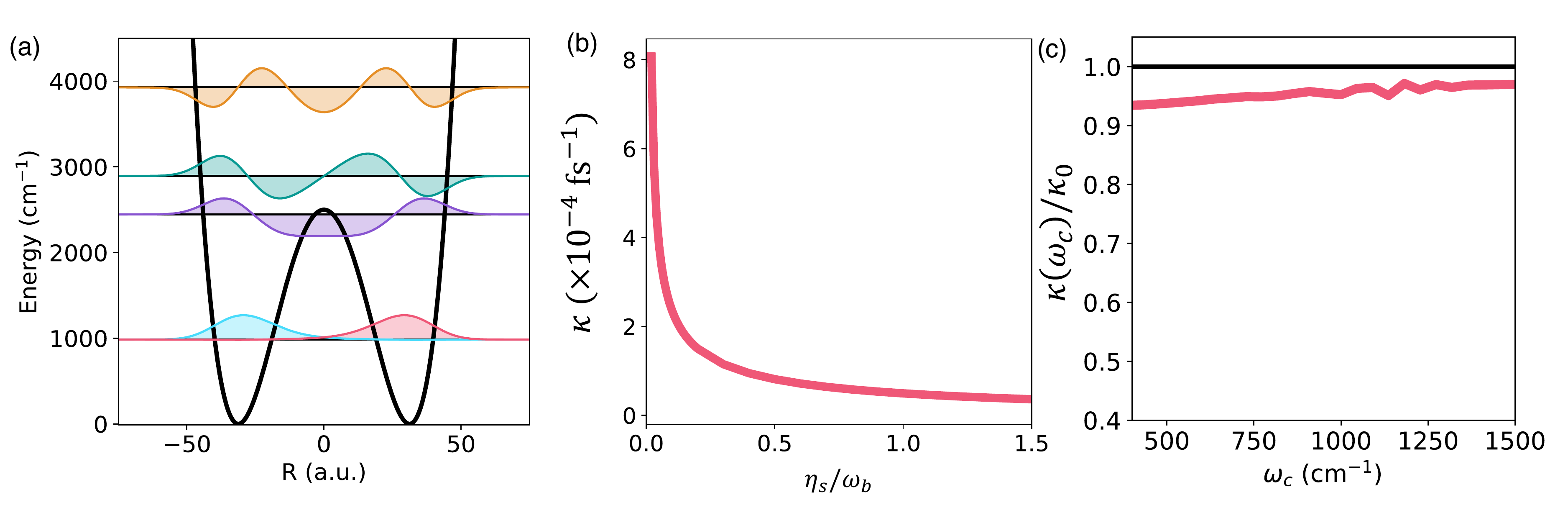}
\caption{ \textbf{Cavity modification of ground state chemical kinetics.} (\textbf{a})   Potential energy surfaces for molecular ground adiabatic state with vibrational eigenstates. (\textbf{b}) Chemical rate constant as function of solvent friction. (\textbf{c}) Cavity modified chemical rate constant as function of cavity photon frequency $\omega_c$.  }
\label{fig2}
\end{figure*}

\subsubsection{Asymmetric Model Potential}

In Fig. \ref{fig3} we present cavity frequency-dependent rate profiles obtained for a model with an unstructured spectral density with friction constant $\eta_s = 0.1\omega_b$ and for an asymmetric reaction coordinate potential with $\omega_b = 1030$ cm$^{-1}$, $E_b = 2730$ cm$^{-1}$, and $c = 0.005$ a.u.  This choice of parameters lead to vibrational transition frequencies in the reactant well that do not differ significantly from those considered in the main text. We have considered a lossy cavity mode with $\tau_c = 1000$ cm$^{-1}$ and $\eta_c = 0.005$ a.u.  In contrast to the results in the main text, here we observe a small shoulder beyond the main peak in the rate profile, this shoulder occurs at a frequency corresponding to the transition between the ground and first excited state of the product well, and can be attributed to a cavity enhancement of the energy loss associated with product well.  It is worth noting that this peak is considerably smaller than that observed for the reactant well, and this can additionally be rationalized in terms of energy loss-based arguments.  In the absence of the cavity,  the energy loss associated with the product well is considerably larger than that associated with the reactant well (owing to the increased depth of the well).  As the rate profile becomes exponentially insensitive to cavity loss in the high cavity loss limit \cite{PollakJCP1989}, the larger starting loss associated with the product well leads to a less significant cavity enhancement.

\begin{figure*}[h]
\centering
\includegraphics[width=0.66\linewidth]{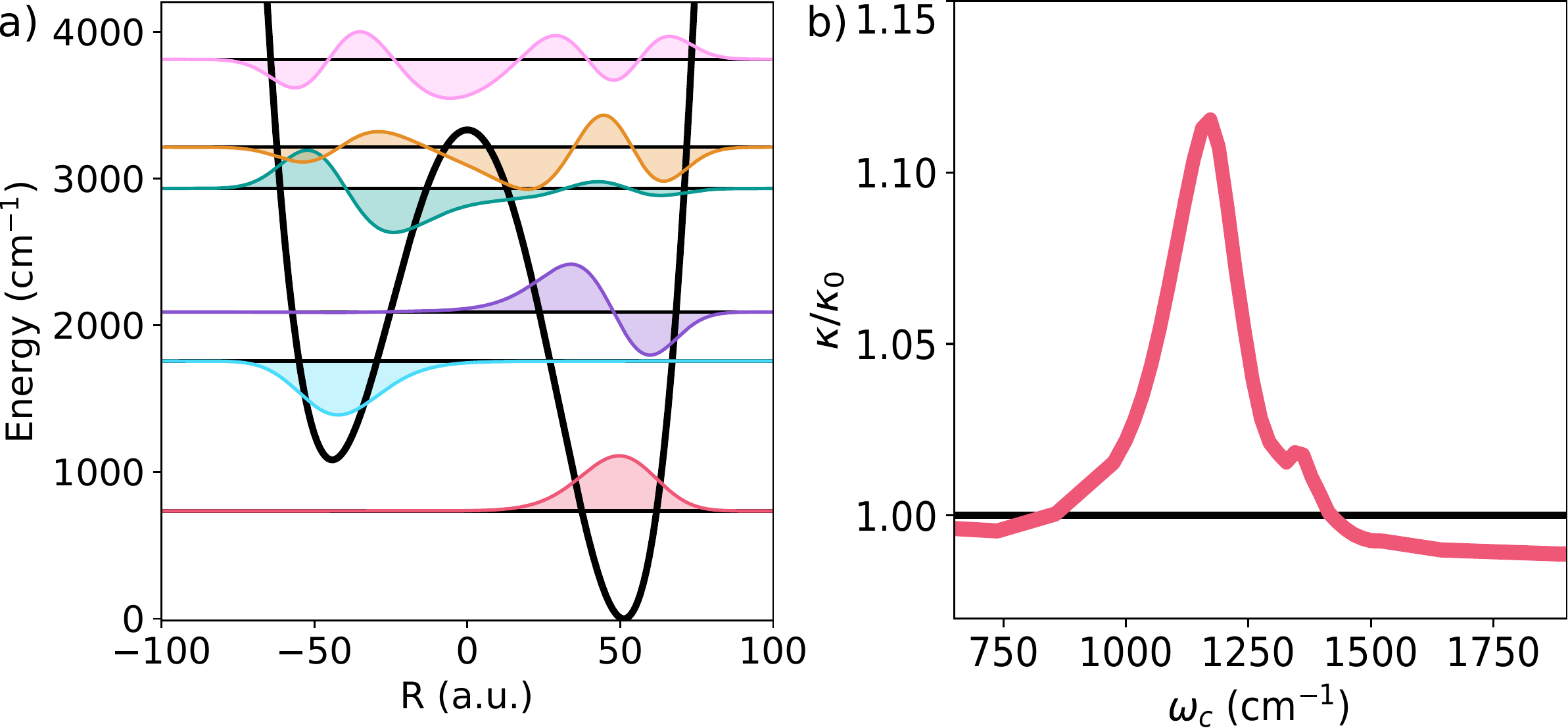}
\caption{ \textbf{Cavity modification of ground state chemical kinetics.} (\textbf{a})   Potential energy surfaces for molecular ground adiabatic state with vibrational eigenstates. (\textbf{b})  Cavity modified chemical rate constant as function of cavity photon frequency $\omega_c$.   }
\label{fig3}
\end{figure*}

\subsection{Additional Structured Spectral Density Results}

\begin{figure*}[h]
\centering
\includegraphics[width=0.66\linewidth]{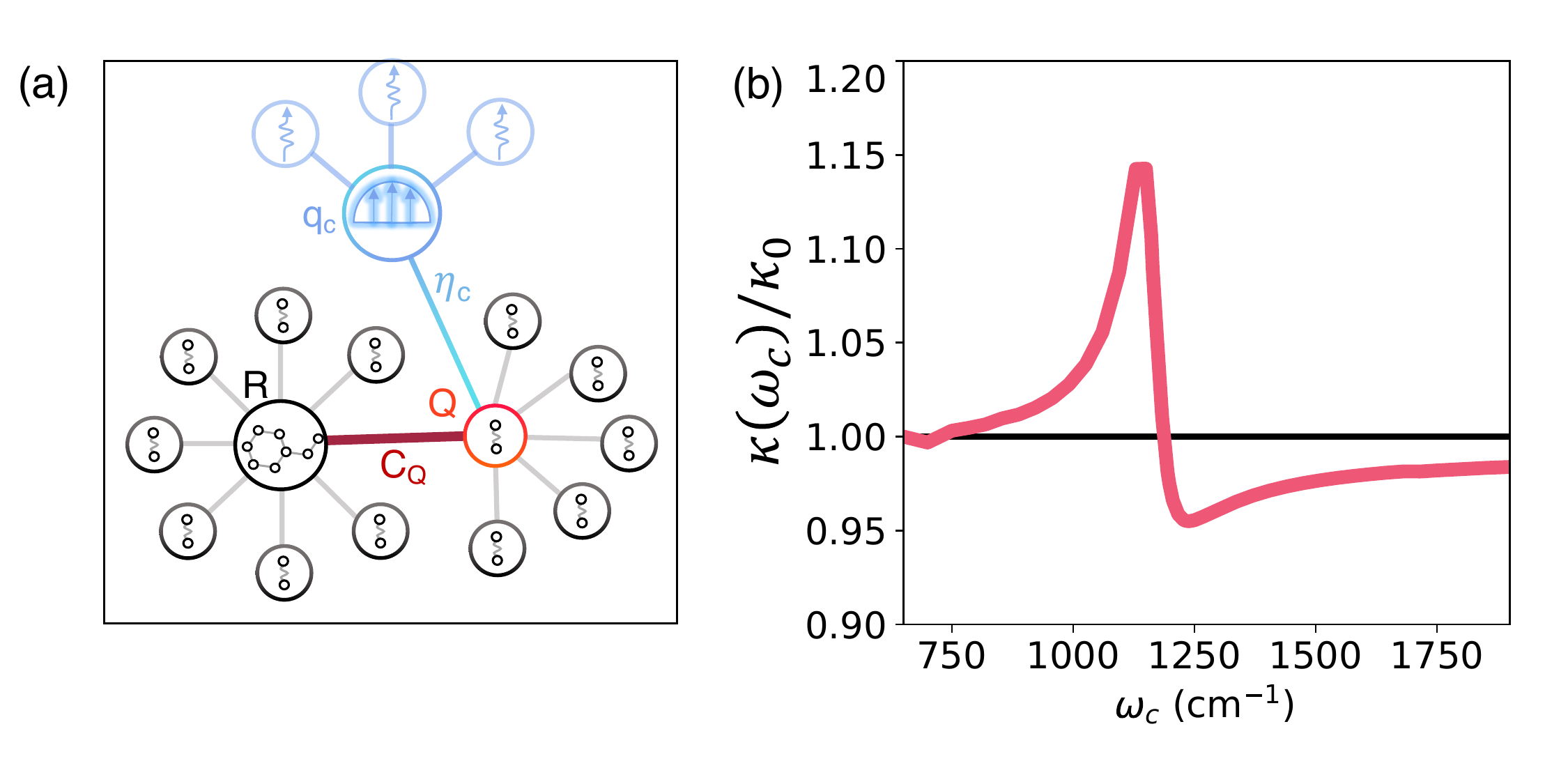}
\caption{ \textbf{Cavity modification of ground state chemical kinetics.} (\textbf{a})   Schematic illustration of a cavity-molecule-solvent setup, with cavity coupling to the spectator mode $Q$. (\textbf{b})   Cavity modified chemical rate constant as function of cavity photon frequency $\omega_c$.  }
\label{fig4}
\end{figure*}

\noindent Finally we consider a system with a structured spectral density arising from the presence of an underdamped spectator mode that couples to the reaction coordinate.  This model was considered in Fig. 4 of the main text, however, here we will consider a lower frequency spectator mode with $\omega_Q = 1100$ cm$^{-1}$.  In Fig. \ref{fig4}, we present the cavity frequency dependent rate profile obtained for this model when a lossy cavity mode, with $\tau_c = 1000$ cm$^{-1}$ is coupled, with $\eta_c = 0.005$ a.u., to the spectactor mode (the case considered in Fig. 4 (d-f) of the main text).  

 \section{Supplementary Note 3: Classical Dynamics}
We write the Hamiltonian as,
\begin{align}
\hat{H} &= \hat{H}_\mathrm{mol} +\hat{H}_\mathrm{solv} + \hat{H}_\mathrm{cav} +   \hat{H}_\mathrm{loss} \nonumber \\
&= \frac{P^2}{2} + V(R) +  \frac{{p}_\mathrm{c}^{2}}{2} + \frac{1}{2} \omega_\mathrm{c}^2 \Big( {q}_\mathrm{c} + \sqrt{2\over \omega_{c}} \eta_c \cdot R  \Big)^2 + \sum_k \frac{{\Pi}_{k}^{2}}{2} + \frac{1}{2}\tilde{\omega}_{k}^{2} \Big({\mathcal{Q}}_k + \frac{\mathcal{C}_{k} {q}_c}{\tilde{\omega}_k^{2}}\Big)^2 + \sum_j \frac{{P}_{j}^2}{2} + \frac{1}{2}\Omega_{j}^2 \Big({X}_j + \frac{C_{j} {R}}{\Omega_j^2}\Big)^2.
\end{align}
We write the equation of motion for $R$ and $q_\mathrm{c}$ as
\begin{align}
\ddot{R}(t) &= -\nabla_{R} V(R) - \omega_\mathrm{c}^2 \Big( {q}_\mathrm{c} + \sqrt{2\over \omega_{c}} \eta_c \cdot R  \Big) \sqrt{2\over \omega_{c}} \eta_c  + \int_0^{t} \dot{R}(\tau) \cdot \eta_{B}(t-\tau) + \xi_{B}(t), \nonumber \\
\ddot{q}_{c}(t) &=   - \omega_\mathrm{c}^2 \Big( {q}_\mathrm{c} + \sqrt{2\over \omega_{c}} \eta_c \cdot R  \Big)  + \int_0^{t} \dot{q}_\mathrm{c}(\tau) \cdot \eta_{c}(t-\tau) + \xi_{c}(t).
\end{align}
The corresponding friction kernel is given as
\begin{align}
\eta_{B}(t) &= \sum_{i} \frac{C_{i}^2}{\Omega_i^2} \cos(\Omega_i t) = \frac{2}{\pi}\int_{0}^{\infty} d\Omega \frac{J_{B}(\Omega)}{\Omega} \cos(\Omega_i t)  = \frac{1}{\pi}\int_{-\infty}^{\infty} d\Omega \frac{J_{B}(\Omega)}{\Omega} \cos(\Omega t), \\
&= \frac{2}{\pi} \Gamma \Lambda_\mathrm{s} \int_{-\infty}^{\infty} d\Omega \frac{\cos(\Omega t) }{\Omega^2 + \Gamma^2} =  2\Lambda_\mathrm{s} e^{-\Gamma t}.
\end{align}
Similarly, we have $\eta_\mathrm{c}(t) = 2\lambda_\mathrm{s} e^{-\gamma t}$.
 \\\\
We rewrite the equation of motion in the extended phase space (adding new degrees of freedom $S_B$ and $S_c$) as~\cite{BaczewskiJCP2013},
\begin{align} 
\dot{P} &=  -\nabla_{R} V(R)  - \omega_\mathrm{c}^2 \Big( {q}_\mathrm{c} + \sqrt{2\over \omega_{c}} \eta_c \cdot R  \Big) \sqrt{2\over \omega_{c}} \eta_c   + S_\mathrm{B}(t), \\
\dot{p}_\mathrm{c} &=  - \omega_\mathrm{c}^2 \Big( {q}_\mathrm{c} + \sqrt{2\over \omega_{c}} \eta_c \cdot R  \Big)  + S_\mathrm{c}(t), \\
\dot{q}_\mathrm{c} &=  {p}_\mathrm{c}, ~~~~~~~~\dot{R} =  P, \nonumber \\
\dot{S}_{B} &= -\Gamma S_{B}(t) - 2\Lambda_s P(t) + 2\Gamma\sqrt{ \Lambda_s / \Gamma \beta } \dot{W}_B(t), \\
\dot{S}_{c} &= -\gamma S_{c}(t) - 2\lambda_s p_c(t) + 2\gamma\sqrt{ \lambda_s / \gamma \beta } \dot{W}_c(t).
\end{align}
The equations of motion are then evolved using a velocity Verlet-type algorithm as provided in Ref.~\cite{BaczewskiJCP2013}. In order to compute the chemical rate, we compute the transmission coefficient from the flux-side correlation function formalism~\cite{FrenkelSmit} as follows
\begin{equation}\label{kappa}
  \kappa (t) = \frac{\langle {\mathcal F}(0) \cdot
  {h}[R(t)-R_{\ddagger}]\rangle}{\langle {\mathcal F}(0)
  \cdot { h}[\dot{R}_{\ddagger}(0)]\rangle},
\end{equation}
where ${h}[R-R_{\ddagger}]$ is a Heaviside function. Here, the dividing surface $R_{\ddagger} = 0$   separates the reactant and the product and $\langle ...\rangle$ 
represents the canonical ensemble average (with constraint on the dividing surface which is enforced by $\delta[R(t)-R_{\ddagger}]$ inside ${\mathcal F}(t)$). Further, $\dot{R}_{\ddagger}(0)$ represents the initial velocity of the reaction coordinate on the dividing surface that is obtained by sampling a classical Maxwell-Boltzmann distribution at $T = 300$ K. The reaction coordinate is initialized at the dividing surface $R(t = 0) = 0$ while the extended variables $S_c$ and $S_B$ are initialized from a gaussian distribution with the distribution widths $\sqrt{2\Lambda_s/\beta \Delta t}$ and $\sqrt{2\lambda_s/\beta \Delta t}$ ( time step $\Delta t = $ 10 a.u.), respectively~\cite{BaczewskiJCP2013}. A total of 600,000 configurations are released from the dividing surface, with the initial momentum $p_c$ and $P$  
randomly sampled from the classical Maxwell-Boltzmann distribution. Each trajectory is propagated for 8 ps such that the flux-side correlation function has plateaued.

\section{Supplementary Note 4: Effects of Cavity Loss on the Rate Constant}

In Fig. \ref{fig5}, we consider properties of the time-dependent transmission coefficient obtained for the symmetric double well potential model considered in the main text with an unstructured spectral density with $\eta_s = 0.1\omega_b$, a light-matter coupling of $\eta_c = 0.00125$ (used to obtain Fig. 3b of the main text).  In Fig. \ref{fig5}a we present the time-dependent transmission coefficients obtained for this model in the absence of cavity loss.  For a cavity frequency of $\omega_c = 1195$ cm$^{-1}$ (near resonant with the reaction coordinate vibration), the transmission coefficient exhibits a significant short-time enhancement.  During this period of time, significant oscillations observed in the instantaneous rate can be attributed to oscillations of the cavity mode.   These oscillations decay with time as the system approaches thermal equilibrium. This decay of oscillations is correlated with a decay of the transmission coefficient, which approaches the results obtained off resonance in the long time limit.  

\begin{figure*}[h]
\centering
\includegraphics[width=\linewidth]{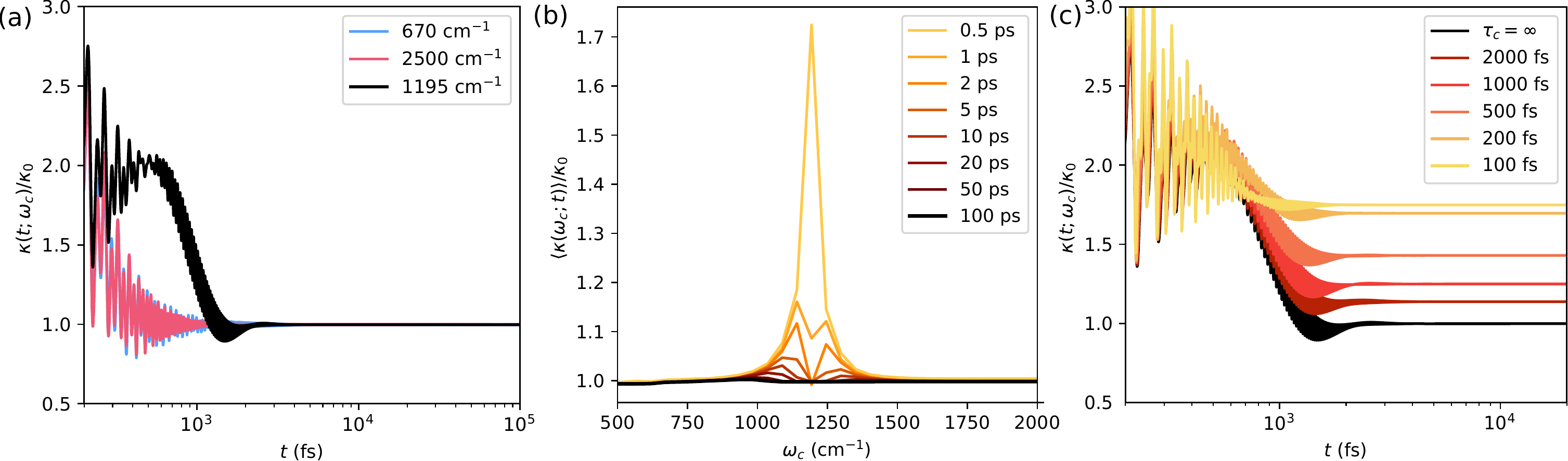}
\caption{\label{fig5}  (a) Time-dependent normalized instantaneous rate constants obtained in the absence of cavity loss for three different values of the cavity frequency.  (b) The average, normalized, instantaneous rate constants obtained over a period of 500 fs starting at different points in time.  (c) The time-dependent instantaneous rate constant obtained for a cavity frequency of $1195$ cm$^{-1}$ and with varying values of the cavity loss.  All results here are obtained for the unstructured spectral density model with $\eta_s = 0.1\omega_b$ considered in the main text and with a light-matter coupling of $\eta_c = 0.00125$. In all cases the instantaneous rate constants are normalized by the out-of-cavity rate.}
\end{figure*}

\noindent The initial enhancement of the instantaneous rate constant, and the decay of this enhancement as the cavity mode thermalizes, is further illustrated in Fig. \ref{fig5}b.   Here the average instantanous rate constant obtained over a period of 500 fs starting at different times is shown as a function of the cavity frequency.  At short times, significant enhancement in the instantaneous rate constant are observed.  At longer times this enhancement decays most rapidly at resonance with the feature splitting into two peaks,  with maxima that occur at larger detuning from resonance at longer times. 
\\\\
Finally, in Fig. \ref{fig5}c, the effect of cavity loss on the time-dependent instantaneous rate constant is shown.  As the cavity loss is increased, the oscillations in the instantaneous rate constant decay more rapidly, indicating more rapid thermalization of the cavity mode.  In addition, this increase in the rate of thermalization of the cavity mode coincides with an increase in the plateau value of the instantaneous rate constant, and therefore an enhancement of the thermal rate constant.  
\\\\
These effect can be understood by considering energy transfer processes between the reaction coordinate - cavity subsystem (containing only $R$ and $q_c$ and described by $\hat{H}_\mathrm{mol} + \hat{H}_\mathrm{cav}$) and its interaction with a dissipative bath composed of the solvent degrees of freedoms $\{X_j\}$ and the far-field cavity modes $\{\mathcal{Q}_{k}\}$. For the molecule-cavity system initially at equilibrium in the left well, the average cavity position will be displaced from equilibrium.  Conversely, if the system was at equilibrium in the right well, the average cavity position will be displaced from equilibrium in the opposite direction. As such, reaction from the left to right will result in non-equilibrium, and generally high energy, configurations for the cavity mode.  This process leads to a loss in energy from the reaction coordinate, which, in the low friction regime, results in an enhancement of the instantaneous rate constant.
\\\\
In the absence of cavity loss, thermalization of the cavity mode will occur through energy transfer from the cavity mode to the molecular bath that is mediated by transfers between the cavity mode and reaction coordinate.  This process depends sensitively on the detuning of the cavity mode and reaction coordinate frequency. For high energy cavity configurations, this process requires a transfer of energy from the cavity to the reaction coordinate decreasing the energy loss of the reaction coordinate, and in turn decreasing the rate constant.
\\\\
Cavity loss provides an alternative pathway for thermalization of the cavity mode, enhancing the rate of thermalization.  Additionally, this pathway does not require transfer of energy from the cavity mode to the reaction coordinate, and enables the cavity mode to act as an additional source of dissipation increasing the reaction rate in this regime.
\\\\
Additionally, these results demonstrate that care should be taken when evaluating cavity-modified chemical reaction rates in the absence of cavity loss.  Depending on the extent of detuning from resonance, the timescales associated with slow energy transfer processes can determine the timescale over which the rate constant plateaus.  

}
\bibliographystyle{naturemag}
\bibliography{bib.bib}